\pdfoutput=1
\documentclass[a4paper,fleqn]{cas-sc}

\usepackage[numbers]{natbib}
\usepackage{booktabs}
\usepackage{amsmath,amssymb}
\usepackage{multirow}
\usepackage{enumitem}
\usepackage{graphicx}
\usepackage[ruled,linesnumbered,vlined]{algorithm2e}
\usepackage{array}
\usepackage{ragged2e}
\usepackage{xcolor}
\graphicspath{{./}{./figures/}}

\newcolumntype{Y}[1]{>{\RaggedRight\arraybackslash}p{#1}}
\newcolumntype{Z}[1]{>{\Centering\arraybackslash}p{#1}}

\providecommand{\num}[1]{#1}
\newcommand{\pct}[1]{\num{#1}\%}
\newcommand{\clvr}{C-LVR}
\newcommand{\lvr}{LVR}
\newcommand{\eetlvr}{E2E-LVR}

\begin{document}
\let\WriteBookmarks\relax
\def\floatpagepagefraction{1}
\def\textpagefraction{.001}

\shorttitle{EndoGov: Knowledge-Governed Multi-Agent EC Risk Stratification}
\shortauthors{Dai et al.}

\title [mode = title]{EndoGov: A knowledge-governed multi-agent expert system for endometrial cancer risk stratification}

\author[1]{Weiye Dai}
\fnmark[1]

\author[2]{Liyun Shi}
\fnmark[1]

\author[4]{Zanxiang He}

\author[1]{Yuling Ma}

\author[2]{Mengyuan Lin}

\author[3]{Dianxiang Sun}

\author[1]{Liming Nie}
\cormark[1]
\ead{nieliming@sztu.edu.cn}

\affiliation[1]{organization={Shenzhen Technology University},
            city={Shenzhen},
            country={China}}

\affiliation[2]{organization={Shenzhen People's Hospital},
            city={Shenzhen},
            country={China}}

\affiliation[3]{organization={Nanyang Technological University},
            city={Singapore},
            country={Singapore}}

\affiliation[4]{organization={Shenzhen University},
            city={Shenzhen},
            country={China}}

\cortext[1]{Corresponding author}
\fntext[1]{Weiye Dai and Liyun Shi contributed equally to this work.}

\begin{abstract}
Multimodal artificial intelligence models for endometrial cancer (EC) risk stratification typically optimize aggregate predictive performance but provide limited mechanisms for enforcing mandatory guideline overrides, such as assigning POLE-mutated tumors to the low-risk group despite high-grade morphology. We present EndoGov, a two-tier multi-agent expert system that factorizes the decision process as $D(x)=G(P(x),\mathcal{R})$, where specialist agents $P$ extract structured evidence and a governance agent $G$ applies an executable rule set $\mathcal{R}$. Tier~1 comprises pathology, molecular, and clinical agents that independently generate schema-constrained reports from frozen foundation-model features or structured records. Tier~2 queries an evidence-level-weighted Guideline Knowledge Graph, using deterministic hard-path rules for high-priority overrides and constrained soft-path reasoning for ambiguous cases. In TCGA-UCEC ($n{=}541$), EndoGov achieved $0.943$ accuracy, $0.973$ macro AUC, and a conditional logic-violation rate (C-LVR) of $0.93\%$ among trigger-exposed cases. In CPTAC-UCEC ($n{=}95$), where reference labels are guideline-derived, EndoGov reached $0.842$ accuracy compared with $<0.31$ for locked-transfer neural baselines, supporting governance-pathway transfer under distribution shift rather than validation against independent clinical truth. End-to-end safety decomposition localized residual failures primarily to upstream molecular detection rather than downstream governance. Backend-swap experiments further showed that hard-path compliance is invariant to the LLM backend. These findings indicate that explicit clinical-rule governance can provide guideline-compliant, auditable EC risk assignment while preserving competitive discrimination.
\end{abstract}

\begin{keywords}
multi-agent systems \sep endometrial cancer \sep knowledge graph \sep clinical decision support \sep guideline governance \sep risk stratification
\end{keywords}

\maketitle

\section{Introduction}
Endometrial cancer (EC) is one of the most common gynecological malignancies, and its management increasingly depends on the joint interpretation of morphology, molecular subtype, and stage. Contemporary EC risk stratification therefore sits at the intersection of pathology, genomics, and clinical staging rather than within any single modality~\citep{kasius2021risk}. Current guidelines also impose an explicit hierarchy over these evidence sources. In the ESMO and ESGO/ESTRO/ESP frameworks, selected molecular findings take precedence over conventional clinicopathological descriptors~\citep{oaknin2022esmo,concin2021esgo}. POLE ultramutation is the clearest example: once confirmed, the case is assigned to the low-risk group irrespective of high-grade morphology~\citep{church2015prognostic,stelloo2016improved}. At the same time, the broader EC literature continues to expand across additional modalities, including multiparametric MRI radiomics~\citep{meng2024mri} and serum biomarkers such as CA-125~\citep{reijnen2019ca125}, underscoring the practical complexity of decision-making in this disease.

This hierarchical clinical structure is poorly aligned with the objective functions used in most multimodal learning pipelines. Representative fusion models, including MCAT, Porpoise, and HECTOR, are trained to improve average discrimination by integrating histomorphology and omics embeddings~\citep{chen2021mcat,chen2022pan,volinsky2024hector}. Such models can achieve strong population-level AUC while still failing in rule-critical subgroups, because dense morphological evidence may dominate a sparse but clinically decisive molecular signal. More broadly, studies comparing human and machine intelligence in EC screening indicate that aggregate performance does not necessarily imply reliability in clinically critical subgroups~\citep{hart2020screening}.

We define this failure mode as a \emph{logic blind spot}: a model may perform well on average while violating a high-priority clinical rule in a minority subgroup for which the appropriate decision is already specified by guidelines. In EC, this problem is particularly consequential because the relevant molecular events are both uncommon and clinically dominant. The central challenge is therefore not only to improve multimodal prediction, but also to structure the decision process so that uncertain perception and explicit rule control coexist within the same deployment pipeline.

\paragraph{Formal decision decomposition.}
An end-to-end model maps input $x$ directly to a decision, $D(x)=f(x)$, leaving no point at which a guideline rule can intervene. A post-hoc rule wrapper, $D(x)=R\bigl(f(x)\bigr)$, applies overrides only after the model has already committed to a latent decision, which limits auditability and prevents rule-guided evidence gathering or confidence calibration. EndoGov instead decomposes the decision as
\begin{equation}\label{eq:decomp}
    D(x) \;=\; G\!\bigl(P(x),\; \mathcal{R}\bigr),
\end{equation}
where $P$ (the specialist agents) converts raw evidence $x$ into structured reports, $\mathcal{R}$ is the executable rule set stored in the Guideline-KG, and $G$ (the governance agent with validator) produces the final decision under the constraints of $\mathcal{R}$. This decomposition makes the system auditable at the component level: $P$ can be evaluated for extraction quality, $\mathcal{R}$ against published guidelines, and $G$ for rule compliance. It also yields a portable systems hypothesis: for another guideline-driven oncology task, the decision architecture can remain fixed while the rule set $\mathcal{R}$ is replaced.

On this basis, we formulate EndoGov as a two-tier multi-agent architecture for inference-time guideline governance. Tier~1 includes pathology, molecular, and clinical agents that operate on frozen features or structured records and generate independent specialist reports. Tier~2 comprises a governance agent that queries the Guideline-KG through coordinated rule-matching and semantic-retrieval channels: the former resolves trigger-dominant cases through deterministic overrides, whereas the latter supplies clause-level evidence for grey-zone reasoning. When no hard trigger is active, the system performs constrained LLM reasoning grounded in retrieved guideline material. The underlying design principle is that separating evidence generation from guideline-governed decision control provides a more appropriate structure for EC risk stratification than optimizing a single fused predictor alone.

The main contributions are:
\begin{enumerate}[leftmargin=*]
    \item \textbf{Problem formulation.} We formalize the logic-blind-spot problem for multimodal EC models and empirically show that standard average-loss multimodal fusion can yield guideline-violating errors in rule-critical subgroups where clinical guidelines mandate deterministic overrides.
    \item \textbf{System architecture.} We propose EndoGov, a two-tier multi-agent architecture for inference-time guideline governance. The decomposition $D(x)=G(P(x),\mathcal{R})$ separates perception from decision control, supports deterministic hard-path execution alongside constrained soft-path reasoning, and generalizes to other guideline-driven oncology tasks by replacing the rule set~$\mathcal{R}$. The contribution is primarily architectural and systems-level rather than algorithmic.
    \item \textbf{Safety-oriented evaluation protocol.} We evaluate governance as a layered property---trigger exposure, governance-layer reliability, end-to-end safety, calibration, and grey-zone behavior---rather than reporting discrimination alone, establishing the first benchmark for four-tier ESMO risk classification on TCGA-UCEC with explicit guideline-compliance metrics.
\end{enumerate}

Three distinctions from common alternatives are worth highlighting. First, unlike post-hoc rule wrapping, EndoGov retrieves rules and validates the decision inside the inference loop rather than after a classifier has committed to a label. Second, unlike open-ended LLM agents, free-form reasoning is confined to the soft path and bounded by retrieved guideline clauses. Third, unlike standard multimodal studies, we treat trigger exposure, governance reliability, and end-to-end safety as separate evaluation targets rather than collapsing them into discrimination alone.

\section{Related work}
\subsection{Foundation models for computational pathology and omics}
Large-scale self-supervised learning has advanced pathology representation learning. UNI, Virchow, TITAN, and related foundation models provide high-quality slide embeddings that support multiple downstream pathology tasks without task-specific end-to-end training \citep{chen2024uni,vorontsov2024virchow,ding2024titan}. In parallel, foundation models for biomedical transcriptomics, such as scGPT, offer transferable molecular encoders that can summarize high-dimensional RNA profiles into compact semantic representations \citep{cui2024scgpt}. EndoGov uses frozen encoders of this type so that model development can focus on decision governance rather than encoder optimization.

EndoGov is not intended to optimize perception encoders. Instead, it freezes the perception front end and introduces a deterministic governance layer downstream, focusing on whether a deployment pipeline can reliably enforce guideline priorities once critical evidence has been extracted.

\subsection{Multimodal fusion for endometrial cancer}
Recent multimodal EC systems increasingly combine histology with molecular or clinical signals rather than relying on unimodal pathology alone. Representative architectures include MCAT and Porpoise, which model cross-modal interactions through co-attention and gated fusion \citep{chen2021mcat,chen2022pan}. A disease-specific reference is HECTOR, a recent endometrial-cancer model reported on large PORTEC-style recurrence-risk cohorts \citep{volinsky2024hector}. These methods demonstrate strong predictive performance, but their objective remains statistical risk minimization. They do not explicitly represent rule priority, cannot naturally produce an audit trail of guideline overrides, and may remain vulnerable to morphology-dominant shortcuts when decisive molecular events are rare. More recent dense-interaction architectures such as SurvPath model biological-pathway-level interactions between histology and genomics for pan-cancer survival prediction~\citep{jaume2024survpath}; although SurvPath targets survival endpoints rather than guideline risk tiers, it represents a state-of-the-art reference for multimodal integration quality. EndoGov is complementary in that it focuses on post-perception governance rather than representation learning.

Beyond histology--genomics fusion, the broader EC risk-prediction landscape is active across multiple modalities. A recent bibliometric review covering 2015--2024 surveys the full spectrum of AI-enabled prognostication approaches in EC, providing a systematic map of the evolving methodological landscape~\citep{wang2026ecbiblio}. Multiparametric MRI radiomics models have achieved promising preoperative risk stratification without tissue sampling~\citep{meng2024mri}, with dedicated radiomic models demonstrating strong discrimination for Stage~I EC risk groups~\citep{chen2021mri} and more recent work linking MRI radiomics signatures to underlying choline metabolic alterations, thereby improving biological interpretability~\citep{lin2024mri}; single-cell and multi-omics integration studies have identified novel stromal-cell-fate gene signatures for UCEC prognostication~\citep{zhang2025stromal}; machine-learning-derived prognostic models based on programmed cell death-related gene signatures offer complementary molecular risk stratification~\citep{chen2025pcd}; immune-response signatures provide additional prognostic value beyond molecular subtyping~\citep{wang2024immuneec}; clinical-grade AI models for EC molecular subtyping from histology alone have been validated across multiple Chinese centers~\citep{qi2025ecmolecular}; and serum CA-125 combined with clinicopathological features improves preoperative risk stratification in low-grade EC~\citep{reijnen2019ca125}. EndoGov differs from these approaches by not introducing a new representation-learning or feature-engineering method. Instead, it treats perception outputs as inputs to a deployment-time governance layer designed to enforce guideline priorities, preserve auditability, and constrain free-form model reasoning. In principle, modality-specific models such as MRI radiomics, gene-signature classifiers, or immune-based scores could be incorporated as additional Tier-1 specialist agents, enriching the evidence packet without altering the governance architecture.

\subsection{Knowledge-enhanced and agentic medical AI}
Knowledge graphs and neuro-symbolic approaches have improved interpretability and consistency in medical AI, while LLM agents have begun to simulate clinical workflows. EndoGov constrains agent roles, externalizes executable knowledge into a guideline graph, and applies deterministic validation so that free-form reasoning cannot override active hard rules. Ontology-plus-Bayesian-network hybrids pursue similar goals of auditability but represent uncertainty through probabilistic graphical structure rather than hard-priority overrides \citep{nzomo2025kgbn}.

LLM-powered agentic systems have opened new avenues for clinical AI. Foundation models encode broad clinical knowledge~\citep{singhal2023large} and serve as biomedical reasoning engines~\citep{tu2024generalist}, while pathology-specific models have scaled to diverse clinical-grade tasks~\citep{moor2023foundation,sun2024cpathomni,yan2025pathorchestra,xu2024gigapath}. Multi-agent frameworks have begun to formalize clinical workflows: AgentClinic~\citep{schmidgall2024agentclinic} evaluates AI agents in simulated clinical environments; MedAgent-Pro~\citep{xu2025medagentpro} introduces an evidence-based agentic workflow for multimodal diagnosis; CLARITY~\citep{shaposhnikov2025clarity} provides triage and routing with explicit reasoning chains; Kg4Diagnosis~\citep{zuo2024kg4diagnosis} enhances multi-agent LLMs with knowledge-graph retrieval; and SmartPath~\citep{xu2025smartpathr1} augments pathology co-pilots with reasoning capabilities. In the computational pathology domain, deep learning methods for tumor grading~\citep{khoshdeli2018deep} and computational pathology surveys~\citep{srinidhi2021deep} laid the foundation for automated histological assessment, while heterogeneous graph transformer methods have been applied to interpretable cancer driver gene prediction~\citep{xiong2023hgtdg}. For endometrial cancer specifically, AI-based histopathology has revealed distinct cancer subsets~\citep{fremond2024ai}, molecular-based prognostic models using miRNA signatures have been developed~\citep{peng2023mirna}, and indicator knowledge graphs have been constructed using retrieval-augmented LLMs~\citep{wang2025indicatorkg}.

Two concurrent KG-grounded multi-agent systems merit direct comparison. Mapis~\citep{zhang2025mapis} employs a knowledge-graph--grounded multi-agent framework for PCOS diagnosis, using agents that traverse a medical KG to gather diagnostic evidence and reach a consensus through iterative discussion. OMGs~\citep{zhang2026omgs} orchestrates multidisciplinary-team agents across the ovarian tumor care continuum, employing consensus-based workflows where multiple specialist agents contribute opinions that are synthesized into a joint recommendation. Both systems share EndoGov's premise that guideline adherence benefits from structured agent decomposition; however, key architectural differences separate them. Mapis relies on LLM-driven KG traversal for evidence gathering without a deterministic rule layer---a hallucinated traversal step can propagate silently. OMGs uses consensus-based agent voting rather than priority-encoded rule arbitration, so there is no structural guarantee that a high-priority guideline override will dominate lower-priority signals. EndoGov differs from both by (i)~separating perception from governance into distinct tiers, (ii)~embedding deterministic hard-path rules with explicit priority encoding so that trigger-covered cases bypass free-form reasoning entirely, (iii)~employing a post-hoc validator that enforces rule--label consistency, and (iv)~introducing quantitative safety metrics (\clvr/\eetlvr) that decompose compliance into trigger exposure, governance reliability, and end-to-end safety---metrics absent from both Mapis and OMGs.

EndoGov complements these designs by embedding deterministic priority rules and a post-hoc validator to guarantee hard-trigger compliance, while externalizing decision authority into a static, auditable guideline graph. More broadly, EvoMDT~\citep{liu2026evomdt} proposes a self-evolving multi-agent system for structured clinical decision-making across multiple cancer types, incorporating adaptive role allocation and iterative consensus refinement. While EvoMDT emphasizes evolutionary adaptability and cross-cancer generality, it does not enforce deterministic hard-path overrides or provide quantitative compliance decomposition (C-LVR/E2E-LVR); EndoGov instead prioritizes verifiable rule adherence at the cost of requiring manual rule compilation for each target guideline.

\subsection{Neuro-symbolic methods, constraint learning, and deployment-side safety}
Several lines of work inject domain constraints into neural networks at training time (DeepProbLog, Logic Tensor Networks, semantic-loss approaches). These methods assume differentiable constraints and sufficient training coverage of constrained subgroups---conditions poorly met in EC, where hard-rule subgroups (e.g., POLE ultramutation) are rare and decisive. EndoGov takes a complementary inference-time approach: rather than encoding constraints into weights, it executes them deterministically in the decision loop.

A related deployment-side family adds runtime admissibility checks, safety layers, or conformal prediction sets around safety-critical decisions. EndoGov belongs to this family but differs in that the constraint layer is the decision authority for hard-trigger cases, not merely a reject option. Conformal coverage on the soft path remains a natural future extension. Similarly, recent work compiles clinical practice guidelines into LLM-traversable decision structures; these systems depend on the LLM for path traversal and lack a deterministic hard path or validator-backed arbitration. EndoGov reserves the LLM exclusively for grey-zone soft-path reasoning while hard-trigger decisions are resolved deterministically. The domain-generalization literature (stain normalization, invariant features) addresses the perception problem rather than governance; EndoGov's CPTAC study asks the narrower question of whether rule-path governance remains stable under locked transfer \citep{otalora2019staining}.

\subsection{Clinical decision support systems}
Traditional CDSS research provides mature rule-modeling formalisms. Arden Syntax encapsulates clinical logic in Medical Logic Modules, and GLIF compiles guidelines into hierarchical task networks. These systems provide reliable rule execution but depend on manually structured inputs---they cannot process whole-slide images or raw omics data. Recent hybrid efforts attach deep-learning perception modules to rule engines, but the interface is typically a serial ``predict then apply rules'' pipeline.

In endometrial cancer specifically, the closest existing CDSS is the ENDORISK Bayesian network, which integrates preoperative clinicopathological variables (histological type, grade, pre-operative imaging, and biomarkers) into a probabilistic graphical model for preoperative risk stratification~\citep{reijnen2020endorisk}. ENDORISK has been externally validated on large population-based cohorts, demonstrating robust calibration and discrimination~\citep{grube2023endorisk}. As a probabilistic CDSS, ENDORISK shares several design goals with EndoGov---explicit structure, interpretability, and preoperative risk assessment---but differs in three respects. First, ENDORISK represents uncertainty through Bayesian conditional probabilities, whereas EndoGov enforces deterministic hard-path overrides for trigger-covered cases and confines probabilistic reasoning to the grey-zone soft path. Second, ENDORISK requires manually curated structured inputs (pre-defined clinical variables), whereas EndoGov accepts raw multimodal data (WSI, RNA-seq, clinical tables) and delegates feature extraction to frozen foundation-model encoders. Third, ENDORISK does not incorporate molecular subtype as a top-level priority override, because it was developed before the ESMO/ESGO molecular-integrated risk tables were widely adopted; EndoGov's rule layer encodes this molecular hierarchy directly. The two approaches are complementary: ENDORISK provides well-calibrated preoperative probability estimates from readily available clinical variables, while EndoGov adds post-perception governance that guarantees compliance with guideline-mandated molecular overrides.

EndoGov can be viewed as a hybrid of a perception front end and a rule-governance back end. The front end, consisting of three specialist agents with frozen foundation-model encoders, performs multimodal feature extraction and structured reporting; the back end performs rule matching, grey-zone reasoning, and compliance validation within a unified decision loop. Compared with traditional CDSS, EndoGov reduces reliance on manually curated structured inputs; compared with end-to-end deep models, it preserves the auditability and determinism of explicit rule execution.

\section{Methodology}
\subsection{Problem setting and logic blind spots}
The target task is four-tier EC risk stratification under ESMO-style criteria with label set $\{\mathrm{L},\mathrm{I},\mathrm{HI},\mathrm{H}\}$. Let $x=(x_{path},x_{mol},x_{cli})$ denote pathology, molecular, and clinical evidence, and let $y$ denote the predicted risk group. In standard end-to-end fusion, the prediction rule $f(x)$ is learned primarily to optimize an average loss over all patients. However, when rare trigger events $E$ correspond to strong clinical overrides, optimizing the average loss can create systematic errors on the trigger-covered subset.

We call these errors logic blind spots. Intuitively, they arise when a sparse but dominant clinical cue is overwhelmed by dense statistical evidence. In EC, a POLE-mutated case with visually high-grade morphology is the canonical example. A purely data-driven fusion model may still predict high risk because morphology occupies a large share of representational capacity, even though the guideline requires low risk.

\subsection{Overview of EndoGov}\label{sec:methodology_overview}
\begin{figure}[pos=h]
\centering
\includegraphics[width=\textwidth]{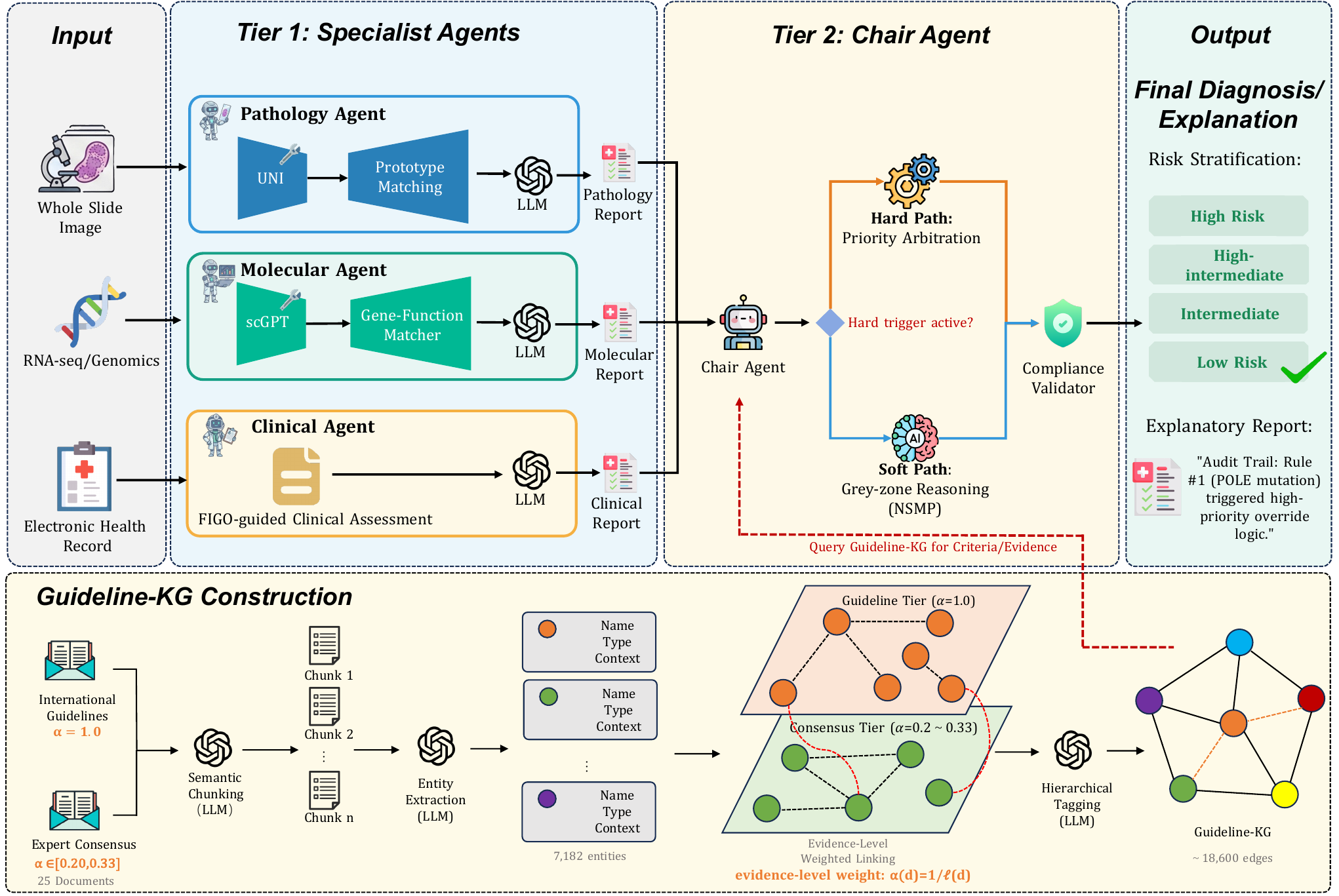}
\caption{Overview of EndoGov (two horizontal bands). \textbf{Upper:} runtime pipeline from multimodal inputs (pathology WSI, molecular omics, structured clinical records) to three Tier~1 specialist agents (UNI$+$prototype pathology matching, scGPT$+$molecular report, and FIGO-guided clinical summarization), then the Tier~2 chair agent, which uses the Guideline-KG and routes each case through hard-path priority arbitration or soft-path grey-zone reasoning, followed by compliance validation to emit a four-tier ESMO-style risk label and an audit-style report. \textbf{Lower:} offline Guideline-KG build from high-weight international guidelines and lower-weight expert-consensus material via semantic chunking, entity/edge construction, evidence-level-weighted inter-tier linking, and hierarchical tagging. Section~\ref{sec:methodology_overview} details the architecture.}
\label{fig:framework}
\end{figure}

Figure~\ref{fig:framework} summarizes the proposed architecture. EndoGov has two tiers:
\begin{enumerate}[leftmargin=*]
    \item \textbf{Tier 1: Specialist agents.} Pathology, molecular, and clinical agents independently summarize modality-specific evidence. This enforces epistemic independence: each modality judges first on its own evidence before the governance layer resolves conflicts.
    \item \textbf{Tier 2: Governance agent.} A chair agent retrieves relevant rules from the Guideline-KG, detects whether a high-priority trigger is active, executes ordered overrides when necessary, and validates the output against the active rule set.
\end{enumerate}

\subsection{Tier 1: role-structured specialist agents}
\paragraph{Dr. Pathology.}
Whole-slide image evidence is encoded with the frozen UNI foundation model \citep{chen2024uni}. Instead of fine-tuning an end-to-end pathology classifier, we use prototype matching in the embedding space to map slide features onto interpretable pathological concepts such as grade, myometrial invasion pattern, and other morphology cues. An LLM then summarizes the matched concepts into a structured pathology report $R_{path}$.

For pathology, let $x_{path}$ denote a WSI patch bag and let $f_{UNI}(\cdot)$ be the frozen UNI encoder. We obtain a slide embedding
\begin{equation}
z_{path} = f_{UNI}(x_{path}) \in \mathbb{R}^{d_p}.
\end{equation}
Given a pathology prototype library $\mathcal{C}_{path}=\{c_1,\dots,c_K\}$, the matcher computes cosine scores
\begin{equation}
s_k = \cos(z_{path}, c_k) = \frac{z_{path}^{\top} c_k}{\|z_{path}\|_2 \|c_k\|_2},
\end{equation}
selects the top-$K$ prototype hits, and maps them to interpretable fields such as histology, grade, and myometrial invasion. The pathology report $R_{path}$ is then a structured summary of these matched concepts rather than a free-form caption.

In the current implementation, these prototypes are not produced by unsupervised $K$-means. Because the target pathology concepts are clinically predefined, we use fold-specific normalized class centroids computed only from the training split. Let $\mathcal{G}^{(f)}_k$ be the set of training patients in fold $f$ whose normalized pathology label maps to one of six pathology concepts: Serous, ClearCell, Undifferentiated, Mixed, and Endometrioid G1/G2/G3. The prototype labels derive from TCGA histology-type annotations rather than downstream four-tier risk labels. The prototypes therefore support structured morphology reporting and do not directly predict risk tiers, avoiding leakage from risk labels into risk prediction. The fold-specific prototype is defined as
\begin{equation}
c^{(f)}_k = \frac{\sum_{i \in \mathcal{G}^{(f)}_k} z_{path,i}}
{\left\|\sum_{i \in \mathcal{G}^{(f)}_k} z_{path,i}\right\|_2}.
\end{equation}

\noindent\textit{Pathology prototype accuracy.} Across the 5-fold cross-validation ($N{=}530$), prototype matching achieves 0.82 top-1 histology-type accuracy and 0.94 top-3 hit rate for the six morphology concepts (Serous, ClearCell, Undifferentiated, Mixed, Endometrioid G1/G2, Endometrioid G3). The main confusion occurs between Endometrioid G2 and G3, consistent with known inter-observer variability in grade assignment. Pathology centroids are highly stable across folds (mean pairwise cosine similarity 0.97--0.98). Because the pathology report feeds into the governance layer as one of three evidence streams and the hard-path decision depends on molecular triggers rather than morphology alone, moderate pathology-prototype errors do not directly propagate to final risk-label violations.

\paragraph{Dr. Molecular.}
The current implementation adopts a DNA-first molecular workflow rather than an RNA-only surrogate. When direct biomarker fields such as \texttt{POLE\_Status}, \texttt{p53\_Status}, or \texttt{MMR\_Status} are available, the molecular agent prioritizes these fields and records \texttt{detection\_source = dna\_direct}. Only when direct molecular evidence is unavailable does the pipeline fall back to scGPT-based RNA embedding, centroid-style prototype matching, and LLM summarization to construct $R_{mol}$. This design is essential for POLE because the clinically relevant event is a DNA-level exonuclease-domain mutation rather than an RNA label. The biological plausibility of the RNA surrogate is supported by TCGA evidence that POLE-ultramutated endometrial tumors form a transcriptomically distinct cluster characterized by ultra-high somatic mutation burden, elevated neoantigen load, and cytotoxic immune-gene upregulation~\citep{tcga2013ucec}; the PORTEC molecular classification further showed that expression-based clustering can recover POLE status with high concordance~\citep{stelloo2016improved}. Accordingly, the transcriptomic branch is interpreted as a surrogate that exploits downstream expression consequences rather than as direct mutation observation~\citep{vangool2015pole,costas2023immunologic,wang2021frequent,li2024polegene}. Because RNA-based POLE inference remains imperfect (the pipeline shows a 13.4-point accuracy decrease under RNA fallback), DNA-direct detection is used whenever available, and DNA-panel inputs can be incorporated without architectural modification.

If direct biomarker fields $b=\{\texttt{POLE},\texttt{MMR},\texttt{p53}\}$ are available, the molecular subtype is assigned by the deterministic priority operator (following TCGA convention, we use \textit{p53abn} throughout to denote the copy-number-high/p53-abnormal subtype, abbreviated CNV-H in some TCGA publications; in clinical practice this corresponds to aberrant p53 immunohistochemistry or \emph{TP53} mutation)
\begin{equation}
\hat m =
\begin{cases}
\text{POLEmut}, & \text{if POLE is mutated},\\
\text{MMRd}, & \text{else if MMR is deficient},\\
\text{p53abn}, & \text{else if p53 is abnormal},\\
\text{NSMP}, & \text{otherwise}.
\end{cases}
\end{equation}
In the deployed decision protocol, p53 abnormality is normalized to the subtype token \texttt{p53abn} and treated as the molecular trigger that activates the high-risk rule family, including the Stage~IA no-myometrial-invasion exception. This representation is deterministic and reproducible, but it also defines a strict validator boundary: if upstream subtype extraction fails to surface \texttt{p53abn}, the validator cannot invoke Rule~R2\_P53 retroactively because no hard trigger has been exposed to the governance layer.
If direct biomarkers are unavailable, RNA-seq features $x_{rna}$ are encoded with frozen scGPT embeddings,
\begin{equation}
z_{mol} = f_{scGPT}(x_{rna}) \in \mathbb{R}^{d_m},
\end{equation}
and matched to molecular prototypes $\mathcal{C}_{mol}=\{\mu_1,\dots,\mu_M\}$ by
\begin{equation}
q_j = \cos(z_{mol}, \mu_j), \qquad \hat m = \arg\max_j q_j.
\end{equation}

For the RNA branch, the molecular prototypes are again fold-specific normalized class centroids rather than unsupervised clusters. Let $\mathcal{H}^{(f)}_j$ denote the training patients in fold $f$ with molecular subtype $j \in \{\text{POLEmut},\text{MMRd},\text{p53abn},\text{NSMP}\}$. Then
\begin{equation}
\mu^{(f)}_j =
\frac{\sum_{i \in \mathcal{H}^{(f)}_j} z_{mol,i}}
{\left\|\sum_{i \in \mathcal{H}^{(f)}_j} z_{mol,i}\right\|_2}.
\end{equation}

\paragraph{Dr. Clinical.}
Structured clinical inputs, including age, stage, grade-related modifiers, and other tabular descriptors, are parsed into a report $R_{cli}$. Rule-based FIGO staging logic is embedded here so that the governance layer receives a clinically coherent summary rather than raw table fields.

\begin{algorithm}[t]
\SetAlgoLined
\caption{Tier-1 Specialist Evidence Generation}
\label{alg:tier1}
\KwIn{pathology input $x_{path}$; direct biomarker fields $b=\{\texttt{POLE},\texttt{MMR},\texttt{p53}\}$; optional RNA profile $x_{rna}$; clinical table $x_{cli}$; frozen encoders $f_{\mathrm{UNI}}, f_{\mathrm{scGPT}}$; fold-specific prototype libraries $\mathcal{C}_{path},\mathcal{C}_{mol}$}
\KwOut{schema-constrained specialist reports $(R_{path},R_{mol},R_{cli})$ with confidence and source metadata}
\BlankLine
\tcc{Pathology branch: morphology evidence, not direct risk prediction}
$z_{path} \leftarrow f_{\mathrm{UNI}}(x_{path})$\;
$H_{path} \leftarrow \mathrm{TopK}_{k}\{\cos(z_{path},c):c\in\mathcal{C}_{path}\}$\;
$\phi_{path} \leftarrow \mathrm{MapToSchema}(H_{path};\{\texttt{histology},\texttt{grade},\texttt{MI}\})$\;
$R_{path} \leftarrow \mathrm{FormatReport}(\phi_{path},\mathrm{confidence}(H_{path}))$\;
\BlankLine
\tcc{Molecular branch: DNA-direct evidence has priority over RNA surrogate inference}
\eIf{$\mathrm{Available}(b)$}{
  $\hat m \leftarrow \mathrm{PrioritySubtype}(b)$ \tcp*{POLEmut $\succ$ MMRd $\succ$ p53abn $\succ$ NSMP}
  $s_{mol} \leftarrow \texttt{dna\_direct}$; $q_{mol} \leftarrow 1.0$\;
}{
  $z_{mol} \leftarrow f_{\mathrm{scGPT}}(x_{rna})$\;
  $j^* \leftarrow \arg\max_j \cos(z_{mol},\mu_j),\ \mu_j\in\mathcal{C}_{mol}$\;
  $\hat m \leftarrow \mathrm{Subtype}(\mu_{j^*})$; $s_{mol} \leftarrow \texttt{rna\_fallback}$; $q_{mol} \leftarrow \cos(z_{mol},\mu_{j^*})$\;
}
$R_{mol} \leftarrow \mathrm{FormatReport}(\texttt{molecular\_subtype}{=}\hat m,\texttt{detection\_source}{=}s_{mol},\texttt{confidence}{=}q_{mol})$\;
\BlankLine
\tcc{Clinical branch: stage and modifier normalization}
$\phi_{cli} \leftarrow \mathrm{ParseClinical}(x_{cli})$\;
$\phi_{cli} \leftarrow \mathrm{NormalizeFIGO}(\phi_{cli})$\;
Derive modifier flags $\{\texttt{LVSI},\texttt{deep\_MI},\texttt{no\_MI}\}$ and consistency checks\;
$R_{cli} \leftarrow \mathrm{FormatReport}(\phi_{cli})$\;
\BlankLine
\Return{$(R_{path}, R_{mol}, R_{cli})$}\;
\end{algorithm}

\subsection{Guideline-KG and priority encoding}
The governance memory that supports both hard- and soft-path reasoning is implemented as a purpose-built Guideline Knowledge Graph (Guideline-KG) for guideline-governed risk stratification. Inspired by the general principle of graph-structured knowledge retrieval demonstrated in medical KG-RAG systems~\citep{wu2025medgraphrag}, EndoGov adapts that paradigm to the present setting through three structural modifications.

First, we deliberately retain only two source tiers---guideline documents and expert consensus documents---and exclude all patient-level data from the evaluation cohorts (TCGA-UCEC and CPTAC-UCEC); the knowledge graph contains only guideline tables and expert consensus documents that exist independently of these cohorts. Including evaluation-cohort cases in the knowledge graph would constitute answer injection, creating a direct information leak from the test set into the retrieval substrate and invalidating any governance evaluation.

Second, we replace unweighted cross-document linking with \textit{evidence-level-weighted linking}, reflecting that not all clinical sources carry equal authority (Section~\ref{sec:elw}).

Third, on top of the MedGraphRAG-derived entity--document graph, we introduce an \textit{executable rule layer} with priority encoding, trigger conditions, and override relationships that is entirely absent from the original framework (Section~\ref{sec:governance}).

\paragraph{Data sources.}
The Guideline-KG is built from a curated corpus of 25 documents organized into two source tiers following MedGraphRAG's Triple Linking strategy:
\begin{itemize}[leftmargin=*]
    \item \textbf{Guideline tier.} 11 international clinical practice guidelines (ESMO 2022, ESGO/ESTRO/ESP 2021, ASTRO 2023, among others) that define the authoritative risk-stratification tables from which executable rules are derived.
    \item \textbf{Consensus tier.} 14 domain-specific expert consensus documents covering molecular subtyping workflows, fertility-sparing management, immunotherapy indications, and regional practice adaptations, which provide supplementary clinical context for grey-zone reasoning.
\end{itemize}
The two tiers total 1.78M characters. Each document is assigned a graph-scoped identifier and annotated with publication metadata (organization, year, source tier).

\paragraph{Graph construction pipeline.}
Graph construction proceeds in four stages.

\textbf{Stage~1: Document chunking.} Source PDFs and HTML are ingested, cleaned, and segmented into chunks. Following MedGraphRAG's semantic chunking strategy, each chunk respects the graph-construction LLM's context limit and is annotated with its source document identifier. In the active build, chunk boundaries are set at 1{,}200 tokens with 100-token overlap.

\textbf{Stage~2: Entity extraction and intra-document linking.} Each chunk is processed by GPT-4o with a structured extraction prompt that identifies medical entities (molecular markers, histological types, staging concepts, clinical features, procedures, and medications) with UMLS-aligned semantic types. Intra-chunk relationships are identified by a second LLM pass, producing directed subgraphs per chunk. The current build yields 7{,}182 entity nodes, each stored with a 1{,}536-dimensional embedding. Quality assurance: (i)~extraction is constrained to seven predefined UMLS-aligned categories; (ii)~a stratified sample of 200 entities was independently reviewed by two researchers, yielding 91\% precision, 87\% recall, and inter-annotator Cohen's $\kappa{=}0.83$; (iii)~near-duplicate entities were merged via embedding-based deduplication (cosine $>0.95$).

\textbf{Stage~3: Evidence-level-weighted cross-document linking.}
\label{sec:elw}
Standard KG-RAG treats all source documents as equally authoritative. EndoGov introduces an \textit{evidence-level weight} $\alpha(d) \in (0,1]$ defined on a five-level ordinal scale: $\alpha(d) = 1/\ell(d)$, where $\ell(d) \in \{1,\dots,5\}$ is the document's assigned evidence level (lower $\ell$ denotes higher authority under our corpus annotation). The 11 guidelines receive $\alpha{=}1.0$; the 14 consensus documents span $\alpha \in \{0.20, 0.25, 0.33\}$. The weighted linking score between entities $e_i \in d_i$ and $e_j \in d_j$ is
\begin{equation}
w(e_i, e_j) = \cos(\mathbf{h}_{e_i}, \mathbf{h}_{e_j}) \cdot \tfrac{1}{2}\bigl(\alpha(d_i) + \alpha(d_j)\bigr),
\label{eq:weighted_ref}
\end{equation}
and a directed \texttt{REFERENCE} edge is created when $w > \delta_r{=}0.6$, yielding ${\sim}12{,}400$ cross-document edges. This formulation ensures high-evidence entity pairs are linked more aggressively while low-evidence pairs require higher raw similarity. At runtime, retrieval uses a fixed top-$K{=}25$ policy with edge weights as relevance scores.

\textbf{Stage~4: Hierarchical tagging and community detection.} Each Meta-MedGraph is summarized with predefined medical tags (following ASTM standards for electronic health record content categories), and an agglomerative hierarchical clustering method with dynamic thresholding groups related subgraphs into a multi-layer tag hierarchy from detailed to broad. This structure enables efficient top-down retrieval at query time, as described in Section~\ref{sec:governance}.

\paragraph{Executable rule layer.}
On top of the MedGraphRAG-derived entity--document graph, EndoGov introduces an additional executable rule layer that is absent from the original framework. A dedicated rule compiler transforms the most authoritative guideline clauses into 12 machine-executable, priority-ranked rules aligned to ESMO 2022 Table~2 and ESGO/ESTRO/ESP 2021 Table~2. Each rule is a first-class graph node (\texttt{ExecutableRule}) connected to typed trigger conditions (\texttt{Trigger}) via \texttt{ACTIVATED\_BY} edges and to a risk outcome (\texttt{RiskOutcome}) via a \texttt{LEADS\_TO} edge. Priority is encoded as an integer, with smaller values indicating stronger clinical dominance: priority~1 for POLE overrides, 2 for p53-abn exceptions, 3--4 for advanced-stage and high-risk-histology overrides, and 10 for soft-path Table~2 inference mappings. \texttt{OVERRIDES} and \texttt{EXCEPTION\_OF} edges encode the dominance hierarchy between rules, and \texttt{DERIVED\_FROM} edges link each rule back to its source clauses. The rule compilation followed a three-step protocol: (i)~the first author translated each qualifying ESMO/ESGO table entry into a structured trigger--outcome--priority tuple; (ii)~a second researcher independently verified each tuple against the source guideline table; (iii)~discrepancies were resolved by returning to the original guideline text. The compiled rules are listed in the released repository; however, this verification protocol involved only two researchers and does not constitute a formal multi-expert validation study with inter-rater agreement statistics. A planned follow-up validation will involve three board-certified gynecologic oncologists independently rating each rule tuple against the cited guideline entries, with Fleiss' $\kappa$ reported for agreement. Table~\ref{tab:rules} summarizes the 12 compiled rules.

\begin{table}[pos=h]
\caption{Summary of 12 compiled executable rules in the Guideline-KG. Priority 1--4: hard-path deterministic overrides; priority 10: soft-path Table~2 inference mappings. ``$\succ$'' denotes the dominance order among triggers.}
\label{tab:rules}
\centering
\small
\begin{tabular*}{\tblwidth}{@{\extracolsep{\fill}}clllc@{}}
\toprule
Rule ID & Pri. & Path & Trigger condition & Risk outcome \\
\midrule
R1\_POLE & 1 & Hard & POLE ultramutation confirmed & Low \\
R2\_P53 & 2 & Hard & p53abn (any stage) & High \\
R2\_P53\_EX & 2 & Hard & p53abn + Stage~IA + no MI & High-Int \\
R3\_MMRd\_ADV & 3 & Hard & MMRd + Stage $\geq$ II & High-Int \\
R5\_ADV & 4 & Hard & Stage $\geq$ III (any subtype) & High \\
R6\_HISTO & 4 & Hard & Serous/Clear-cell/Undiff. & High \\
R7\_IVB & 4 & Hard & Stage IVB (any subtype) & High \\
S1\_NSMP\_LO & 10 & Soft & NSMP + IA + G1-2 endo & Low \\
S2\_NSMP\_INT & 10 & Soft & NSMP + IB or G3 & Intermediate \\
S3\_MMRd\_INT & 10 & Soft & MMRd + IA/IB & Intermediate \\
S4\_MMRd\_HI & 10 & Soft & MMRd + LVSI+ or deep MI & High-Int \\
S5\_DEFAULT & 10 & Soft & No other rule matches & Chair reasoning \\
\bottomrule
\end{tabular*}
\end{table}

Formally, the governance memory is a heterogeneous graph
\begin{equation}
G=(V,E), \qquad V=V_D \cup V_R \cup V_E,
\end{equation}
where $V_D$ are document/chunk nodes (from Stage~1), $V_R$ are executable rule, trigger, and risk-outcome nodes (from the rule layer), and $V_E$ are entity/community nodes (from Stage~2). The edge set $E$ contains ${\sim}12{,}400$ evidence-level-weighted cross-document \texttt{REFERENCE} edges (Stage~3, Eq.~\ref{eq:weighted_ref}), ${\sim}5{,}800$ intra-document \texttt{CONTAINS}/\texttt{HAS\_ENTITY} edges, and 48 rule-layer edges (\texttt{ACTIVATED\_BY}, \texttt{LEADS\_TO}, \texttt{OVERRIDES}, \texttt{EXCEPTION\_OF}, \texttt{DERIVED\_FROM}), totaling approximately 18{,}600 edges.

\begin{figure}[pos=h]
\centering
\includegraphics[width=\textwidth]{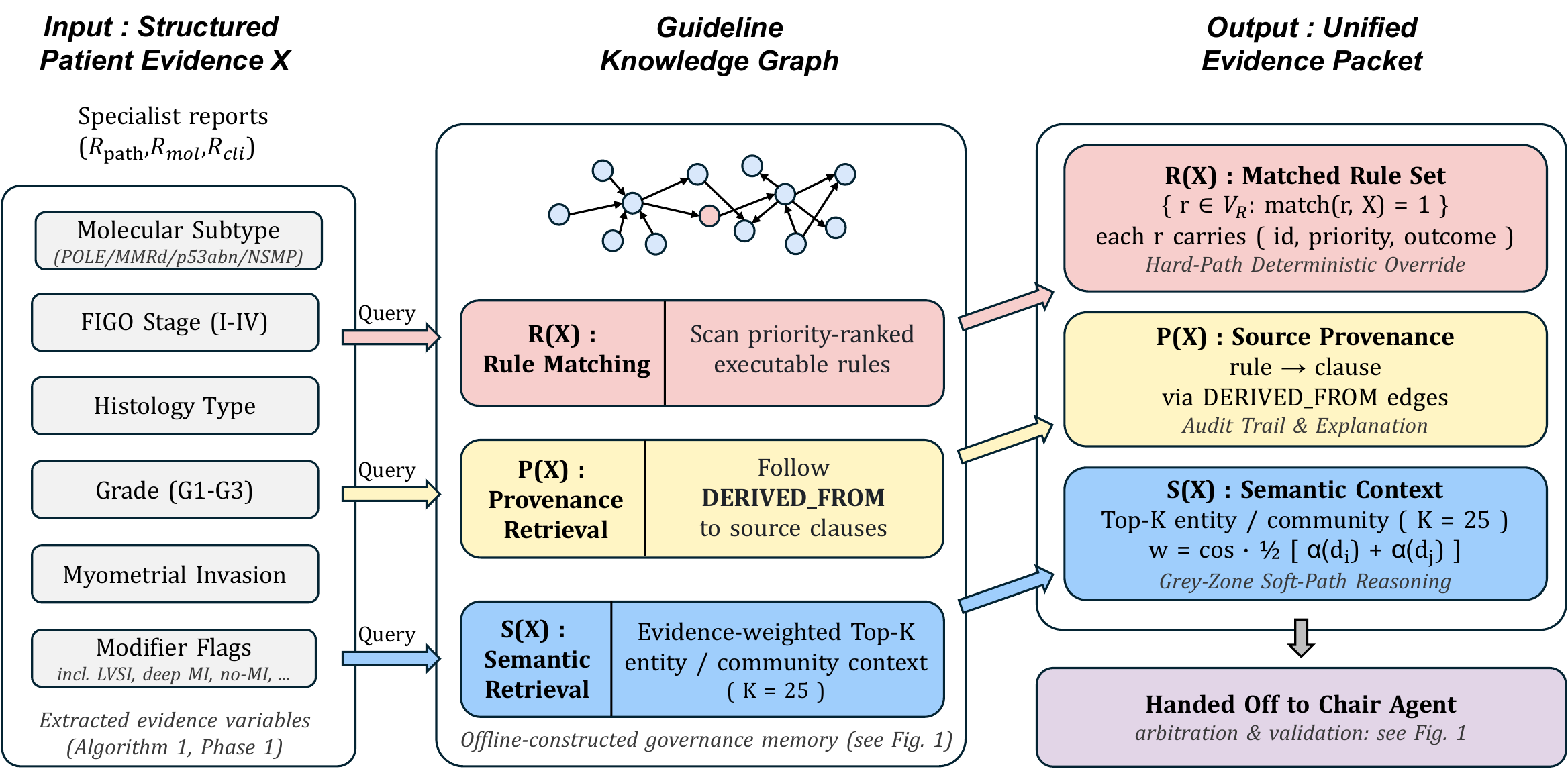}
\caption{Runtime contract over the Guideline-KG. \textbf{Input} (left): structured patient evidence $X$ is extracted from specialist reports ($R_{path}$, $R_{mol}$, $R_{cli}$) into clinically typed variables, including molecular subtype, FIGO stage, histology, grade, myometrial invasion, and modifier flags such as LVSI, deep myometrial invasion, and no-myometrial-invasion status. \textbf{Graph query} (center): the offline governance memory is queried through three aligned channels---$\mathcal{R}(X)$ scans priority-ranked executable rules, $\mathcal{P}(X)$ follows \texttt{DERIVED\_FROM} edges to source guideline clauses, and $\mathcal{S}(X)$ retrieves evidence-weighted Top-$K$ entity/community context ($K{=}25$). \textbf{Output} (right): the unified evidence packet contains the matched rule set for hard-path deterministic override, source provenance for audit trail and explanation, and semantic context for grey-zone soft-path reasoning; the packet is then handed to the chair agent for arbitration and validation.}
\label{fig:kg_query}
\end{figure}

\paragraph{Online query contract.}
\label{sec:governance}
Given extracted patient evidence $X$, the runtime contract proceeds in three parallel retrieval channels, illustrated in Figure~\ref{fig:kg_query}.

\noindent First, the rule match set is computed:
\begin{equation}
\mathcal{R}(X)=\{r \in V_R : \mathrm{match}(r,X)=1\}.
\end{equation}
If $\mathcal{R}(X)\neq\varnothing$, the deterministic risk proposal is produced by the highest-priority matched rule,
\begin{equation}
r^*=\arg\min_{r\in\mathcal{R}(X)} p(r), \qquad y_{hard}=\mathrm{risk}(r^*).
\end{equation}
Second, source provenance is retrieved by following \texttt{DERIVED\_FROM} edges from matched rules to their source guideline chunks:
\begin{equation}
\mathcal{P}(X)=\{d \in V_D : \exists r \in \mathcal{R}(X),\ (r \xrightarrow{\texttt{DERIVED\_FROM}} d)\}.
\end{equation}
Third, contextual entity evidence is retrieved from the entity layer using the evidence-level-weighted graph (Eq.~\ref{eq:weighted_ref}):
\begin{equation}
\mathcal{S}(X)=\mathrm{Top}\mbox{-}K_{\mathrm{retr}}(X;V_E,w).
\end{equation}
The chair agent receives $\mathcal{R}(X)$ for deterministic inference, $\mathcal{P}(X)$ for source provenance, and $\mathcal{S}(X)$ for contextual reasoning support. Rule priority is not used to manipulate semantic retrieval ranking; rule matching and semantic retrieval are executed in parallel and fused only at the arbitration stage, which avoids converting the retrieval subsystem into a hidden priority-weighted rule engine. All external reruns and confidence intervals reported in this study were generated using the same locked evaluation protocol.

This design occupies an intermediate position between generic KG-RAG and conventional if--else rule engines. Compared with MedGraphRAG, EndoGov introduces three domain-specific adaptations: (i)~a two-tier source design that excludes patient data to prevent answer injection; (ii)~evidence-level-weighted cross-document linking (Eq.~\ref{eq:weighted_ref}) that preferentially retrieves high-authority guideline evidence over lower-level consensus material; and (iii)~an executable rule layer with priority encoding, trigger conditions, and override relationships, enabling deterministic and auditable decisions in hard-path cases rather than retrieval-generated answers. This KG-centric governance design shares the motivation of automated KG-construction methods for risk propagation in safety-critical domains, but differs by encoding explicit priority-ranked clinical rules with deterministic override semantics. Relative to a static rule engine, it preserves an evidence-level-aware entity knowledge base for soft-path grey-zone reasoning while linking each rule to its source through \texttt{DERIVED\_FROM} edges. Rule content is not used to rescore semantic retrieval; instead, the chair consumes a unified evidence packet containing matched rules and retrieved entity context, and the validator performs final consistency checking.

\subsection{Governance agent and arbitration protocol}
Given $R_{path}$, $R_{mol}$, and $R_{cli}$, the chair agent executes a four-step protocol inside the inference loop rather than after a classifier has already committed to a label (Algorithm~\ref{alg:governance}):
\begin{enumerate}[leftmargin=*]
    \item \textbf{Evidence extraction.} Convert the specialist reports into structured evidence variables such as molecular subtype, FIGO stage, histology, grade, myometrial invasion, and modifier flags.
    \item \textbf{Guideline retrieval.} Query the Guideline-KG for rule candidates and supporting clauses relevant to the extracted evidence, together with priority and provenance metadata.
    \item \textbf{Priority arbitration.} If a hard trigger is active, apply the corresponding ordered rule directly, for example POLE $\rightarrow$ Low Risk or p53abn $\rightarrow$ High Risk.
    \item \textbf{Compliance validation.} Check the provisional decision against the active rules; if a contradiction is detected, emit a corrected result together with an audit message.
\end{enumerate}

When no deterministic rule applies, the system enters a soft path where the knowledge graph becomes the central reasoning substrate rather than merely a provenance store. Retrieved entity context and cross-document relationships from the Guideline-KG are passed to the chair agent, which integrates stage, grade, LVSI, histology, and molecular status into a coherent risk assessment grounded in guideline evidence. This is especially important for NSMP and MMRd patients whose final assignment depends on interacting clinicopathological factors rather than on a single dominant trigger. The critical boundary is that the LLM does not have authority over hard-path outcomes: once an active hard rule is matched, the final label is determined by the rule engine, not by free-form generation.

\begin{algorithm}[t]
\SetAlgoLined
\caption{EndoGov Governance Workflow}
\label{alg:governance}
\KwIn{specialist reports $(R_{path},R_{mol},R_{cli})$; Guideline-KG $G=(V_D,V_R,V_E,E)$; executable rules $\mathcal{R}$ with priority $p(r)$ and path tag; validator $V$; semantic retrieval size $K$; grey-zone threshold $\tau$}
\KwOut{final risk label $\hat{y}$, confidence $\hat c$, and audit trail $A$}
\BlankLine
\tcc{Phase 1: schema binding and parallel KG query}
$X \leftarrow \mathrm{ExtractEvidence}(R_{path}, R_{mol}, R_{cli})$\;
$\mathrm{AssertSchema}(X;\{\texttt{subtype},\texttt{stage},\texttt{histology},\texttt{grade},\texttt{MI},\texttt{LVSI}\})$\;
$\mathcal{R}(X) \leftarrow \{r\in V_R:\mathrm{match}(r,X)=1\}$\;
$\mathcal{P}(X) \leftarrow \{d\in V_D:\exists r\in\mathcal{R}(X),(r\xrightarrow{\texttt{DERIVED\_FROM}}d)\}$\;
$\mathcal{S}(X) \leftarrow \mathrm{TopK}_{K}(X;V_E,w)$\;
$M \leftarrow (\mathcal{R}(X),\mathcal{P}(X),\mathcal{S}(X))$\;
$\mathcal{H}(X) \leftarrow \{r\in\mathcal{R}(X):\mathrm{path}(r)=\mathrm{Hard}\}$\;
\BlankLine
\tcc{Phase 2: priority arbitration}
\eIf{$\mathcal{H}(X) \neq \varnothing$}{
  $r^* \leftarrow \mathrm{ResolvePriority}(\mathcal{H}(X),p,\texttt{OVERRIDES},\texttt{EXCEPTION\_OF})$\;
  $y \leftarrow \mathrm{risk}(r^*)$; $c \leftarrow 1.0$\;
  $\pi \leftarrow \mathrm{Trace}(\texttt{hard\_path},r^*,\mathcal{P}(X))$\;
}{
  $(y,c,r_s) \leftarrow \mathrm{MapESMOTable2}(X,\mathcal{R}(X))$\;
  $\pi \leftarrow \mathrm{Trace}(\texttt{soft\_table2},r_s,\mathcal{P}(X))$\;
  \If{$c < \tau$}{
    $(y,c,\pi) \leftarrow \mathrm{ChairReason}(X,M,\pi)$\;
  }
}
\BlankLine
\tcc{Phase 3: validator-bounded output}
$(\hat{y},\hat c,A) \leftarrow V.\mathrm{ValidateAndCorrect}(y,c,X,M,\pi)$\;
\BlankLine
\Return{$(\hat{y},\hat c,A)$}\;
\end{algorithm}

\noindent\textit{Implementation note.} Retrieval and validation are internal operations of the chair workflow rather than separate graph nodes. Soft-path cases are first mapped by deterministic ESMO Table~2 logic; an additional chair-agent reasoning step is used only for lower-confidence grey-zone cases. The chair receives a unified evidence packet from the Guideline-KG rather than a duplicated rule list, reducing repeated prior amplification across retrieval, prompting, and validation. The validator recomputes the guideline-expected risk from schema-constrained evidence fields, compares it with the proposed final label, and accepts, rejects, or corrects the output according to the active matched rules. The retained audit packet includes the matched rule identifiers, decision path, molecular detection source, and validator message; soft-path reasoning is bounded to retrieved clause and entity context. The validator can block contradictions against surfaced rules, but it cannot repair missed upstream triggers, incorrect evidence extraction, or all soft-path hallucination modes when the relevant rule was never activated.

\noindent\textit{Prompt structure overview.} Each specialist agent receives a structured system prompt defining its role, output schema, and evidence-formatting instructions. The pathology prompt specifies fields for histology type, grade, myometrial invasion, and confidence, and maps prototype-matched concepts into these fields without free-form elaboration. The molecular prompt enforces subtype, detection-source, and confidence fields and encodes the deterministic priority hierarchy POLE $\succ$ MMRd $\succ$ p53abn $\succ$ NSMP. The clinical prompt parses tabular inputs into FIGO-consistent staging fields. The chair prompt receives the unified evidence packet, including matched rules $\mathcal{R}(X)$ with priority and provenance metadata, source clauses $\mathcal{P}(X)$, and Top-$K$ entity context $\mathcal{S}(X)$. It then checks whether a hard trigger is active, applies the highest-priority rule when present, otherwise reasons over retrieved guideline clauses to select a risk tier, and returns a structured decision with rule identifier, confidence score, and reasoning trace. The validator receives the proposed label and active matched rules, flags contradictions with the highest-priority rule, and emits a corrected label with an explicit audit reason when necessary. All prompts use temperature${=}0$ and constrained output schemas to reduce stochastic variation.

\subsection{Safety-oriented evaluation metrics}
To quantify compliance, we evaluate logic violations on the subset for which a high-priority rule should be decisive. Let $c_i \in \{0,1\}$ indicate whether patient $i$ triggers a hard rule and let $y_{rule,i}$ denote the risk group mandated by the guideline. The logic-violation rate is
\begin{equation}
\lvr = \frac{1}{N_{trigger}} \sum_{i=1}^{N} \mathbb{I}(c_i=1)\cdot \mathbb{I}(y_i \neq y_{rule,i}),
\end{equation}
where $N_{trigger}=\sum_i \mathbb{I}(c_i=1)$.

Because the governance layer is only expected to guarantee compliance once the upstream trigger has been correctly exposed, we also report the conditional logic-violation rate (\clvr). To separate governance reliability from trigger-recognition quality, we further define the end-to-end logic-violation rate,
\begin{equation}
\eetlvr = \frac{N_{missed\ trigger\ \&\ wrong\ final}}{N_{trigger}},
\end{equation}
which counts trigger-covered cases that fail before they can benefit from deterministic governance. In other words, C-LVR measures the internal reliability of the governance layer conditional on correct trigger exposure, whereas E2E-LVR measures the residual safety burden of the full pipeline. We also report trigger sensitivity, trigger specificity, calibration error, and subgroup-specific AUC.

\noindent\textit{Denominator convention.} The \clvr\ values in Tables~\ref{tab:main_results} and \ref{tab:backend} use the 224-case trigger-exposed subset of the downstream benchmark; the 0.0\% governance-layer \clvr\ in Table~\ref{tab:e2e_safety} uses the separate 212-case adjudicated dominant-trigger audit.

\noindent\textit{ECE caliber note.} The ECE values reported in Table~\ref{tab:main_results} are four-class multiclass ECE, measuring probability calibration across all four risk tiers; they should not be compared directly with binary high-risk-vs.-non-high-risk calibration analyses.

\noindent\textit{C-LVR denominator summary.} Three denominators are used for logic-violation metrics across the paper. To avoid confusion, we consolidate them below.

\smallskip
\noindent\begin{tabular}{@{}p{0.22\linewidth}ccp{0.38\linewidth}@{}}
\toprule
\textbf{Metric} & $\boldsymbol{N}$ & \textbf{Value} & \textbf{Denominator source} \\
\midrule
C-LVR (Tables~\ref{tab:main_results},~\ref{tab:backend}) & 224 & 0.93\% & Runtime trigger-exposed cases \\
Gov.-layer C-LVR (Table~\ref{tab:e2e_safety}) & 212 & 0.0\% & Adjudicated dominant-trigger audit \\
E2E-LVR (Table~\ref{tab:e2e_safety}) & 212 & 2.4\% & Same 212-case audit, incl.\ upstream misses \\
\bottomrule
\end{tabular}
\smallskip

\noindent The 12-case gap (224 $-$ 212) consists of borderline-trigger cases excluded from adjudication. The two violations producing the 0.93\% downstream C-LVR are p53abn patients for whom the Stage~IA no-myometrial-invasion exception was incorrectly applied in the deterministic mapping layer---a rule-boundary error rather than an LLM reasoning failure, identically reproduced across all LLM backends. These two cases fall within the 12-case gap and are therefore excluded from the 212-case adjudicated audit, explaining why governance-layer C-LVR is 0.0\% in Table~\ref{tab:e2e_safety}.

\subsection{Porting to new institutions or diseases}
Porting EndoGov to a new guideline-driven task requires four steps. \textbf{(1)~Rule compilation:} a domain expert translates the target guideline table into priority-ranked trigger--outcome tuples using the released rule schema (Rule~ID, Priority, Path, Trigger, Outcome, Source). \textbf{(2)~Specialist-agent adaptation:} the specialist agents use structured JSON schemas to report modality-specific evidence; new diseases require defining appropriate schema fields and, if the modality changes, swapping the frozen encoder. \textbf{(3)~KG construction:} the MedGraphRAG pipeline is re-executed on the new guideline corpus; the deploying institution assigns ordinal evidence levels $\ell(d)$ to each source document (Section~\ref{sec:elw}). \textbf{(4)~Validation:} \clvr/\eetlvr\ metrics are computed on a labelled pilot cohort to verify hard-path compliance. The estimated effort is 2--4 weeks for a team familiar with the framework.

\section{Experimental design}
\subsection{Analysis cohorts and clinical endpoint}
The primary internal benchmark is the aligned TCGA-UCEC downstream cohort ($N{=}541$), obtained by patient-level linkage of molecular labels, structured clinical records, and WSI assets. The principal denominator chain is as follows: aligned downstream cohort ($N{=}541$) $\rightarrow$ PanCanAtlas-labelled held-out cohort ($N{=}530$) and molecular audit subset ($N{=}529$) $\rightarrow$ runtime trigger-exposed subset ($N{=}224$, used for downstream \clvr) $\rightarrow$ adjudicated dominant-trigger audit ($N{=}212$, used for governance-layer \clvr\ and \eetlvr). The 12-case difference between 224 and 212 comprises borderline-trigger cases excluded from adjudication. Molecular status is reported under two complementary accounting schemes: ground-truth subtypes (POLEmut 48, MMRd 147, p53abn 158, NSMP 188; total 541) and runtime detection-source pathways (365 DNA-direct and 176 RNA-fallback cases). For external assessment, we use an aligned CPTAC-UCEC cohort ($N{=}95$)~\citep{dou2020proteogenomic}; because CPTAC reference labels are guideline-derived, this experiment evaluates governance transfer and rule-path consistency rather than full multimodal perception generalization. The primary endpoint is the ESMO-style four-tier risk group.

\subsection{Feature protocol and implementation}
All neural baselines are retrained under the same frozen-feature protocol used by EndoGov. TransMME uses all three branches, whereas MCAT and Porpoise remain bimodal. External baselines are trained on the internal frozen-feature cohort ($N{=}527$) and transferred to CPTAC without adaptation, so the external comparison is a locked governance-transfer stress test rather than a full multimodal generalization benchmark.

Centroid construction uses fold-specific normalized class centroids computed from training patients only, avoiding information leakage. The chair agent uses a deterministic hard path for trigger-covered cases and constrained LLM reasoning (temperature${}=0$) for grey-zone cases; while temperature${}=0$ is ``mostly deterministic'' rather than strictly so, repeated runs on a stratified 50-case subsample (covering hard-path, soft-path, and all molecular subtypes) produced identical outputs across three independent executions, and all reported results use a single fixed run. The main backend is GPT-4o~\citep{openai2024gpt4o}; backend-swap experiments in Table~\ref{tab:backend} replace only the chair agent with Llama-3.3-70B~\citep{grattafiori2024llama3} and Qwen2.5-72B~\citep{yang2024qwen25}. For trigger auditing, the canonical subtype is recovered from merged biomarker fields (POLE $\rightarrow$ POLEmut, MMR $\rightarrow$ MMRd, p53 $\rightarrow$ p53abn, otherwise NSMP).

\subsection{Guideline-KG utility and boundary analysis}
Because the Guideline-KG is designed for decision support rather than standalone link prediction, we evaluate it using task-relevant utility rather than generic graph metrics alone. Four quantities are most informative: (i) hard-path coverage, quantifying how often the rule channel resolves a case deterministically; (ii) trigger exposure quality, measuring whether clinically dominant triggers are surfaced before arbitration; (iii) governance consistency after exposure, assessing the internal reliability of the matched-rule channel; and (iv) soft-path utility, reflected by the non-trivial fraction of cases requiring entity and cross-document evidence beyond deterministic overrides.

Under this protocol, the Guideline-KG is actively used on both branches of inference rather than functioning only as a passive provenance store. Among the 541 locked TCGA cases, 279 terminate on the deterministic hard path and 262 enter the soft path, confirming that the semantic entity-retrieval channel remains operationally relevant rather than redundant. On the adjudicated dominant-trigger audit subset, trigger sensitivity reaches 94.8\% and trigger specificity reaches 96.1\%, indicating that the rule channel usually exposes the right cases for deterministic governance. Conditional on correct trigger exposure, governance-layer \clvr\ falls to 0.0\%, whereas the end-to-end \eetlvr\ remains 2.4\% because residual failures are dominated by upstream misses and ambiguous evidence rather than by downstream rule arbitration itself.

These observations also clarify the fairness boundary between rule injection and retrieval. In EndoGov, executable rules constrain decision arbitration, but they do not rewrite the evidence-level-weighted relevance function used to retrieve entity evidence. The retrieval channel therefore remains an evidence-graded recall mechanism, while the rule channel acts only after evidence aggregation. This separation reduces the risk that performance gains are caused by repeated injection of the same rule prior into graph ranking, prompting, and validation simultaneously. A minimal retrieval ablation that collapses the richer entity context to executable-rule retrieval alone drops internal accuracy to 0.641 and macro AUC to 0.848 while keeping \clvr\ at 0.0\%, indicating that the evidence-level-weighted entity graph primarily improves grey-zone discrimination rather than hard-path rule compliance. The ablation impact is strongly subgroup-dependent: hard-path subtypes (POLE, p53abn) retain perfect accuracy (1.000), whereas soft-path subtypes deteriorate substantially (MMRd: $0.587$; NSMP: $0.254$), confirming that the entity and cross-document layers are operationally essential for grey-zone reasoning. Future work should systematically evaluate entity-link thresholds, retrieval top-$K$, and chunk-size parameters to characterize their effects on grey-zone performance, particularly NSMP AUC and soft-path calibration.

\subsection{Baselines and statistical protocol}
We compare EndoGov with representative unimodal and multimodal baselines: ABMIL, HIPT, SNN, Porpoise, MCAT, and TransMME \citep{ilse2018attention,chen2022scaling,chen2022pan,chen2021mcat,chen2023transmme}. All rows are reproduced under a unified evaluation protocol rather than copied from prior papers; ABMIL, MCAT, Porpoise, and SNN are simplified reference implementations, whereas TransMME is the strongest full neural baseline in the current benchmark. More recent multimodal fusion architectures---SurvPath~\citep{jaume2024survpath} (CVPR 2024), MOTCat~\citep{xu2023motcat} (ICCV 2023), and the HECTOR fusion backbone~\citep{volinsky2024hector} (Nature Medicine 2024)---are designed and benchmarked for survival-outcome endpoints (c-index) on different cohorts, not for the discrete four-tier ESMO risk classification task evaluated here; to our knowledge, no published work has applied any of them to this specific task on TCGA-UCEC. We therefore discuss them as state-of-the-art references for multimodal integration quality (Section~2.2) without including mismatched task numbers in Table~\ref{tab:main_results}. The governance controls in Table~\ref{tab:backend}, together with the compact iso-information control reported in Section~5.5, provide the more informative fairness and ablation analyses.

For the CPTAC study, baseline heads are trained once on the locked internal standardized-feature cohort with complete frozen-feature coverage ($N=527$) and then evaluated on the aligned external CPTAC cohort without CPTAC-specific fine-tuning. We also include ablations that remove role decomposition or governance components, enabling the contributions of structured agents, executable rules, and the chair loop to be assessed separately.

\section{Results}
\subsection{Overall comparison on TCGA-UCEC}
Table~\ref{tab:main_results} summarizes the internal evaluation on the full aligned cohort ($n=541$). EndoGov shows strong predictive performance while maintaining a low governance-layer \clvr\ (0.93\%) on the 224 trigger-exposed cases defined within the downstream benchmark itself.\footnote{Two trigger denominators are used throughout: $N_{\mathrm{trigger\text{-}exposed}}{=}224$ for downstream \clvr\ and $N_{\mathrm{dominant\text{-}trigger}}{=}212$ for the safety decomposition in Table~\ref{tab:e2e_safety}; see Section~3.5 for definitions.} All baseline rows were rerun on this same cohort under the locked evaluation protocol, so the comparison is patient-aligned rather than cross-paper.

\begin{table}[pos=h]
\caption{Internal benchmark on aligned TCGA-UCEC ($N=541$; \clvr\ uses the downstream 224-case trigger-exposed denominator). The molecular agent achieves 497/529 correct subtype assignments (macro sensitivity 0.945); of 212 adjudicated dominant triggers, 201 are correctly detected and 207 are finally classified correctly (Table~\ref{tab:e2e_safety}).}
\label{tab:main_results}
\centering
\small
\begin{tabular*}{\tblwidth}{@{\extracolsep{\fill}}Y{0.27\linewidth}Z{0.10\linewidth}Z{0.10\linewidth}Z{0.11\linewidth}Z{0.12\linewidth}Z{0.13\linewidth}@{}}
\toprule
Method & Modality & Acc. & Macro AUC & ECE$\downarrow$ & \clvr$\downarrow$ \\
\midrule
ABMIL & WSI & \num{0.781} & \num{0.795} & \num{0.188} & N/A \\
HIPT & WSI & \num{0.812} & \num{0.826} & \num{0.162} & N/A \\
SNN & Mol & \num{0.846} & \num{0.861} & \num{0.149} & N/A \\
Porpoise & WSI+Mol & \num{0.886} & \num{0.898} & \num{0.132} & \pct{7.1} \\
TransMME & All & \num{0.904} & \num{0.908} & \num{0.121} & \pct{6.4} \\
MCAT & WSI+Mol & \num{0.893} & \num{0.901} & \num{0.128} & \pct{7.8} \\
\textbf{EndoGov} & All & \textbf{\num{0.943}} & \textbf{\num{0.973}} & \textbf{\num{0.075}} & \textbf{\pct{0.93}} \\
\bottomrule
\end{tabular*}
\end{table}

\noindent\textit{Bootstrap 95\% CI for EndoGov (2000 resamples):} accuracy $[0.922, 0.963]$, macro AUC $[0.956, 0.987]$, macro F1 $[0.893, 0.947]$. Pairwise DeLong comparison against TransMME (the strongest neural baseline) yields $\Delta$AUC $= +0.065$ ($p{<}0.001$)~\citep{delong1988comparing}; a paired bootstrap test confirms $\Delta$Acc $= +0.039$ ($p{<}0.01$).

Together with the trigger-level audits below, these results show that EndoGov combines competitive cohort-level discrimination with explicit trigger-exposure analysis and near-zero governance-layer failure after correct exposure.

\subsection{Upstream molecular audit and end-to-end safety decomposition}
Because hard-rule enforcement depends on trigger recognition, comprehensive safety reporting should examine the upstream molecular module rather than only the final governed output. On the PanCanAtlas-labelled cohort ($N=529$), the molecular agent produces 497 correct subtype assignments, corresponding to macro sensitivity 0.945 and macro F1 0.943. For dominant trigger cases specifically, the independent audit in Table~\ref{tab:e2e_safety} shows that 201/212 POLE or p53abn triggers were correctly exposed to the hard path and 207/212 trigger cases were finally classified correctly, indicating that the remaining safety burden is concentrated in upstream misses rather than in downstream rule execution after exposure.

On the PanCanAtlas-labelled held-out cohort ($N{=}530$), subtype-level performance is stable rather than driven by a single majority subgroup: POLE reaches 0.943 accuracy ($N{=}35$), MMRd 0.972 ($N{=}142$), p53abn 0.975 ($N{=}162$), and NSMP 0.948 ($N{=}191$), with overall held-out accuracy 0.962. The DNA-first versus RNA-fallback split is also non-trivial: when direct biomarker evidence is available, the downstream governed pipeline reaches 0.986 accuracy on 365 runtime DNA-direct cases; when the molecular subtype must be inferred from RNA alone, accuracy drops to 0.852 on 176 RNA-fallback cases, a 13.4-point absolute decrease. These subtype and detection-source results are retained here as compact textual checks rather than as a separate main-text table.

\begin{table}[pos=h]
\caption{Trigger-case safety decomposition on the adjudicated dominant-trigger audit ($N=212$; governance-layer \clvr\ here is distinct from Table~\ref{tab:main_results}' downstream 224-case denominator).}
\label{tab:e2e_safety}
\centering
\small
\begin{tabular*}{\columnwidth}{@{\extracolsep{\fill}}Y{0.70\columnwidth}Z{0.24\columnwidth}@{}}
\toprule
Quantity & Value \\
\midrule
Ground-truth trigger cases ($N_{trigger}$) & 212 \\
Correctly detected triggers & 201 \\
Trigger sensitivity & \pct{94.8} \\
Trigger specificity & \pct{96.1} \\
Missed triggers & 11 \\
Recovered by soft path despite missed trigger & 6 \\
Final correct classifications among trigger cases & 207 \\
False triggers & 8 \\
Governance-layer \clvr & \pct{0.0} \\
End-to-end \eetlvr & \pct{2.4} \\
\bottomrule
\end{tabular*}
\end{table}

\begin{figure}[pos=h]
\centering
\includegraphics[width=\textwidth]{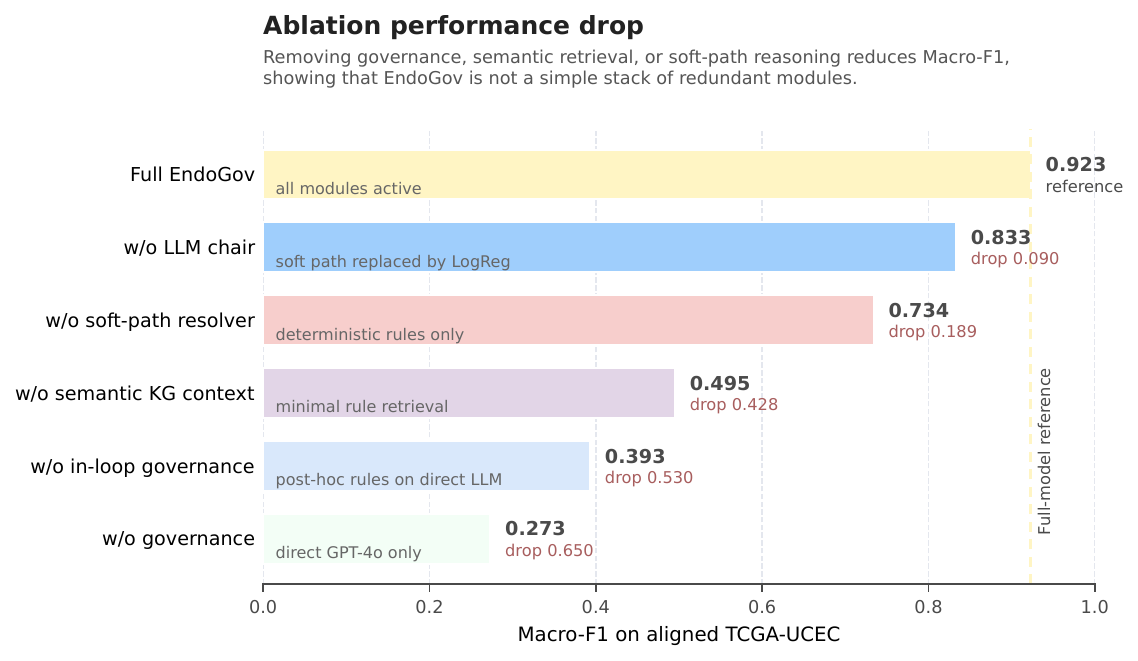}
\caption{Ablation performance drop on aligned TCGA-UCEC. Bars show Macro-F1 for the full EndoGov pipeline and component-removal controls; the dashed vertical line marks the full-model reference (Macro-F1 $=0.923$). Replacing the LLM chair with a logistic-regression soft-path resolver produces a moderate drop, removing the soft-path resolver or semantic KG context produces larger degradation, and replacing in-loop governance with direct/post-hoc LLM variants collapses Macro-F1. This pattern argues against a redundant engineering stack: governance, semantic retrieval, and bounded soft-path reasoning each contribute to final performance.}
\label{fig:ablation_drop}
\end{figure}

This decomposition clarifies the clinical role of governance. Once a hard trigger is surfaced, EndoGov deterministically executes the corresponding guideline rule. Residual errors are therefore attributable primarily to failed trigger exposure, ambiguous upstream molecular evidence, or difficult mixed-evidence cases. To assess whether the architecture contains redundant components, Figure~\ref{fig:ablation_drop} reports a component-removal analysis.

\subsection{External validation and reliability analysis}
The external analysis evaluates whether the governance stack transfers under a reproducible CPTAC-aligned structured-input protocol. Because this experiment does not rerun a complete external WSI-transcription pipeline and uses locked-transfer reference baselines without CPTAC-specific adaptation, it should be interpreted as governance-transfer validation rather than as an external state-of-the-art leaderboard. Table~\ref{tab:external} reports the results.

\begin{table}[pos=h]
\caption{External governance-transfer validation on aligned CPTAC-UCEC ($N{=}95$). Locked transfer from the internal cohort; no CPTAC-specific fine-tuning. Neural baselines: range across ABMIL, SNN, MCAT, Porpoise, and TransMME locked-transfer heads trained on the internal cohort without CPTAC adaptation.}
\label{tab:external}
\centering
\small
\begin{tabular*}{\tblwidth}{@{\extracolsep{\fill}}Y{0.22\linewidth}Y{0.14\linewidth}Z{0.06\linewidth}Z{0.12\linewidth}Z{0.12\linewidth}@{}}
\toprule
Method & Modality & $N$ & Acc. & Macro AUC \\
\midrule
Neural baselines & Frozen & 95 & 0.211--0.305 & 0.460--0.555 \\
Soft-path LogReg & Structured & 95 & \num{0.684} & \num{0.829} \\
\textbf{EndoGov} & Governed & 95 & \textbf{\num{0.842}} & \textbf{\num{0.909}} \\
\bottomrule
\end{tabular*}
\end{table}

\noindent\textit{Bootstrap 95\% CI for EndoGov on CPTAC (2000 resamples):} accuracy $[0.768, 0.916]$, macro AUC $[0.871, 0.945]$, macro F1 $[0.656, 0.873]$. The relatively wide F1 interval reflects the inherent statistical uncertainty of $N{=}95$.

The primary external comparison is between the strongest structured comparator and the full governed system. Neural reference models transferred under the same CPTAC protocol reach 0.21--0.31 accuracy and 0.46--0.56 macro AUC (Table~\ref{tab:external}, stress-test row). This decline is consistent with frozen-feature distribution shift between TCGA and CPTAC, including differences in tissue processing, scanner characteristics, and molecular assay platforms, together with the absence of domain adaptation in the locked-transfer protocol. EndoGov is less sensitive to this shift because hard-path decisions rely on structured biomarker fields, which transfer more reliably across cohorts than learned embedding distributions. Relative to the structured comparator, the soft-path logistic-regression resolver reaches 0.684 accuracy and 0.829 macro AUC, whereas full EndoGov maintains 0.842 accuracy and 0.909 macro AUC. Because CPTAC reference labels are guideline-derived, this protocol should be interpreted as evaluating rule-consistent transfer rather than independent clinical truth.

\paragraph{External error distribution on CPTAC.}
Among the 15 external errors, 14 are high-risk under-calls (class accuracy $13/27=0.481$); low-risk ($51/51$) and high-intermediate ($5/5$) cases are classified correctly. Hard-path cases achieve $17/17$ accuracy, whereas soft-path cases achieve $63/78$ accuracy (0.808). All 15 errors occur among the 20 patients whose detected subtype differs from the CPTAC molecular reference, with the greatest concentration in p53abn cases ($5/20$ correct). This pattern is consistent with the validator boundary: once p53abn is surfaced, Rule R2\_P53 enforces high-risk logic, whereas errors remain uncorrectable when the upstream pathway fails to expose the p53abn trigger.

\paragraph{Conservative referral simulation.}
As a deployment-oriented mitigation, we simulated conservative referral policies. On TCGA, restricting autonomous release to DNA-direct cases and referring all RNA-fallback or unknown cases yields 66.9\% coverage with 0.986 accuracy (5 errors). On CPTAC, releasing only confidence-1.0 hard-path cases and referring all sub-unity-confidence cases yields 17.9\% coverage with 1.000 accuracy, routing all 15 external errors, including all 14 high-risk under-calls, into the referred group.

\paragraph{Scope and limitations of external validation.}
Three design choices shape the interpretation of the CPTAC results. First, the locked-transfer protocol was selected to isolate governance robustness from domain-adaptation effects: by freezing all neural heads and transferring without CPTAC-specific tuning, the experiment evaluates whether the rule layer and structured-input pathway remain stable under distribution shift, rather than whether the complete multimodal stack achieves optimal external generalization. With modest domain adaptation, such as fine-tuning classification heads on a small labelled CPTAC subset or applying stain normalization to WSI features, both neural baselines and EndoGov's soft-path performance would likely improve. The more relevant question is whether the hard-path compliance advantage persists; the backend-swap experiments in Table~\ref{tab:backend} support this interpretation because hard-path \clvr\ is invariant to the soft-path engine. Second, the low locked-transfer accuracy of neural baselines (0.21--0.31) should be interpreted as a stress test rather than as a fair external benchmark for those architectures under standard transfer learning. Third, because CPTAC reference labels are derived from the same ESMO/ESGO guideline tables encoded in EndoGov's rule layer, the external evaluation is partially circular for hard-path cases. This circularity does not invalidate the experiment, which confirms faithful implementation of the intended guideline logic and transferability of structured biomarker fields, but it limits the conclusion to \emph{governance-pathway transfer validation}. Establishing clinical generalizability will require prospective multicenter validation with independently clinician-adjudicated labels across institutions using different molecular assay platforms and pathology workflows; a collaboration with two additional clinical centers has been initiated for this purpose.

\subsection{Mechanistic case analyses}
\begin{figure}[pos=h]
\centering
\includegraphics[width=\textwidth]{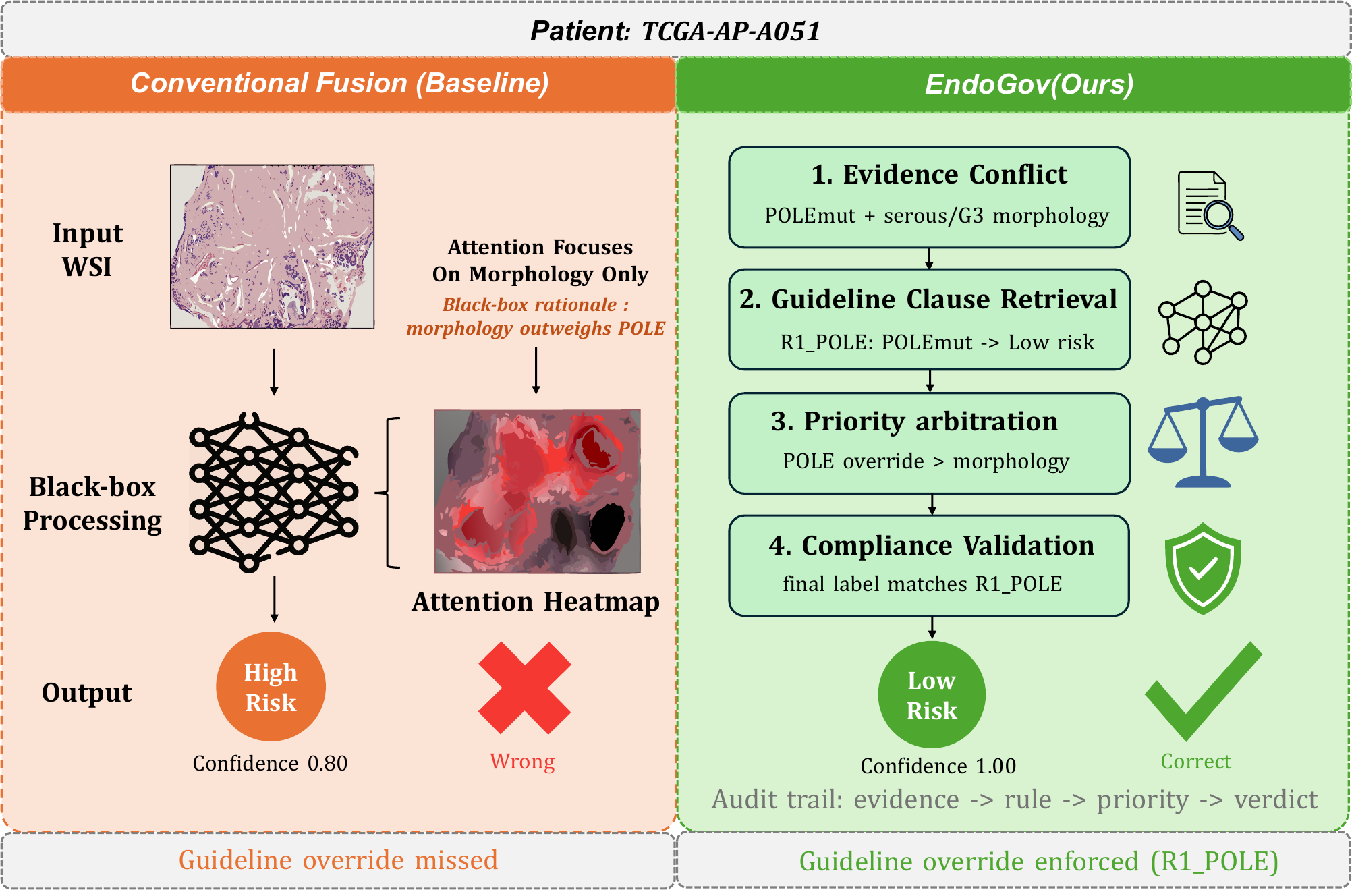}
\caption{Guideline-grounded reasoning chain for an atypical POLE-mutated case (TCGA-AP-A051). \textbf{Left:} a conventional fusion baseline treats high-grade serous morphology as dominant evidence and predicts \emph{High risk}, missing the POLE override. \textbf{Right:} EndoGov parses the conflicting evidence, retrieves the source-linked R1\_POLE clause, applies priority arbitration rather than averaging evidence streams, and validates the final \emph{Low risk} label against the matched rule. The case illustrates how EndoGov converts a black-box failure into a step-by-step, verifiable audit trail.}
\label{fig:mechanism}
\end{figure}

Figure~\ref{fig:mechanism} illustrates the mechanism behind the hard-path safety gain. The example patient (TCGA-AP-A051) is POLE-mutated but presents misleading high-grade serous morphology. A conventional fusion baseline predicts high risk by overweighting morphology and clinical high-risk cues. EndoGov retrieves the relevant POLE rule, recognizes it as a priority-1 hard-path override, and returns the guideline-consistent low-risk label together with a transparent audit trail.

The same framework is also informative in grey-zone cases. For example, an NSMP patient (TCGA-AX-A05Z) with Grade 2 disease, FIGO Stage IB, and negative LVSI is predicted as only intermediate risk by the black-box baseline, which underweights the risk-upgrading role of deep myometrial invasion. In contrast, the chair agent filters the retrieved guideline clauses against the verified evidence, rejects the superficially similar low-risk clause, and combines the remaining admissible rules with structured specialist reports to reach the correct high-intermediate-risk assignment.

To illustrate the soft-path reasoning chain, we trace patient TCGA-D1-A17N (Stage~IA, G2, endometrioid, NSMP). The pathology agent reports ``High-Intermediate'' (confidence 0.5) due to an ambiguous G3-prototype match, while the molecular agent confirms NSMP (``Intermediate'', confidence 0.6) and the clinical agent reports ``Low'' (confidence 0.9). No hard triggers match, so the case enters the soft path. The chair agent retrieves 20 guideline snippets and applies ESMO 2022 Table~2: NSMP + Stage~IA + G1/G2 endometrioid $\rightarrow$ Low risk (matching ground truth). This case highlights how governance anchors the decision to the guideline clause rather than blending conflicting morphological and molecular signals.

\paragraph{Validator intervention statistics.}\label{sec:validator_stats}
Across the 541-case internal cohort, the validator records 261 active interventions or consistency-enforcement events (48.2\%). This high rate is \emph{by design} rather than an indicator of poor upstream reasoning: on hard-path cases, the validator's primary role is to \emph{execute} deterministic rule overrides (e.g., enforcing POLE $\rightarrow$ Low or p53abn $\rightarrow$ High), not to correct LLM errors. Of the 279 hard-path cases, all received deterministic override rules (R1\_POLE: 49, R2\_P53: 166, R5\_ADV: 54, R6\_HISTO: 5, R7\_IVB: 5), and the remaining 262 cases entered ESMO Table~2 mapping and optional chair-agent reasoning, where the validator checked the proposed label against the matched soft-path rule set. The intervention breakdown is: 166 (63.6\%) for p53-abnormality rule enforcement, 54 (20.7\%) for advanced-stage overrides, 33 (12.6\%) for POLE and histology overrides (R1\_POLE: 28 label rewrites + R6\_HISTO: 5), and 8 (3.1\%) for soft-path staging inconsistencies. Thus, most validator activity is rule execution and consistency enforcement, not free-form error correction. The validator cannot generate new evidence or repair missing triggers; its operational scope is bounded by the schema-constrained fields surfaced by the specialist agents.

\paragraph{Soft-path failure analysis.}
Among the 262 soft-path cases, 22 are misclassified (soft-path accuracy 0.916). All 22 errors involve NSMP patients (22/143 NSMP soft-path cases; MMRd soft-path accuracy is 1.000 on 119 cases). The dominant misclassification patterns are: Low $\rightarrow$ Intermediate (9 cases), Low $\rightarrow$ High-Intermediate (7 cases), Intermediate $\rightarrow$ High-Intermediate (3 cases), High-Intermediate $\rightarrow$ Intermediate (2 cases), and High-Intermediate $\rightarrow$ High (1 case). The systematic upward bias (16 of 22 errors are over-classifications) is consistent with the conservative principle but results in non-trivial false-alarm burden for genuinely low-risk NSMP patients. All 22 errors originate from RNA-inference detection paths; none occur in DNA-direct soft-path cases, reinforcing the finding that the RNA surrogate is the dominant error source in the grey zone.

\subsection{Subgroup analysis and supplementary controls}
Subgroup-level reporting clarifies where discrimination remains a perception problem and where rule-based governance largely collapses uncertainty. In the current 541-case cohort, POLE and p53abn are near-single-label subgroups after guideline mapping, so macro AUC is not informative there; MMRd and especially NSMP remain the more relevant subgroups for discrimination analysis. The pattern is clinically consistent: POLE and p53abn are largely resolved by deterministic guideline priority once correctly exposed, whereas NSMP remains the subgroup in which multimodal evidence integration still materially determines the final label. In our locked cohort, NSMP remains the hardest subgroup but still reaches 0.853 accuracy and 0.944 macro AUC under the full governed pipeline.\footnote{NSMP AUC in Table~\ref{tab:backend} denotes the macro AUC computed only on patients with NSMP molecular subtype, evaluating four-tier risk classification within this subgroup. Because NSMP lacks a deterministic hard rule and risk depends on the combination of multiple clinicopathological factors, this metric specifically measures the system's grey-zone discrimination capacity.}

\paragraph{Baseline architecture coverage.}
As noted in Section~4.4, the baselines span 2018--2023; more recent architectures (SurvPath, MOTCat, HECTOR) target survival endpoints rather than four-tier ESMO classification, making direct comparison infeasible. As an iso-information control, we also tested variants in which baselines receive an additional 4-dimensional molecular-subtype one-hot feature. Even the strongest gated-fusion variants (MCAT-Gated: 0.706 accuracy, 3.3\% LVR; Porpoise-Gated: 0.708 accuracy, 1.4\% LVR) remain below EndoGov and cannot guarantee hard-trigger compliance, confirming that EndoGov's near-zero governance-layer \clvr\ is an architectural property rather than a consequence of baseline vintage.

\subsection{Direct-LLM and cross-backend validation}
The central architectural claim of this paper is not that GPT-4o alone solves the task, but that structured agents and governance change what the system can guarantee. Because reviewer-facing controls are especially important for separating role decomposition, governance, and backend effects, Table~\ref{tab:backend} reports the corresponding supplementary experiments under the same 541-case cohort, identical specialist evidence, fixed prompts, zero-temperature decoding, and the same evaluation protocol.

\begin{table}[pos=h]
\caption{Governance and backend controls on TCGA-UCEC. \clvr\ is reported on the downstream 224-case trigger-exposed subset.}
\label{tab:backend}
\centering
\small
\begin{tabular*}{\tblwidth}{@{\extracolsep{\fill}}Y{0.24\linewidth}Y{0.18\linewidth}Z{0.08\linewidth}Z{0.09\linewidth}Z{0.08\linewidth}Z{0.08\linewidth}Z{0.11\linewidth}@{}}
\toprule
Setting & Control & Acc. & Macro AUC & ECE & \clvr & NSMP AUC \\
\midrule
\multicolumn{7}{@{}l}{\textit{Pure symbolic controls}} \\
\textbf{Structured Rules Only} & Deterministic ESMO Table~2 on structured reports & \num{0.800} & \num{0.828} & \num{0.200} & \pct{0.93} & \num{0.747} \\
\textbf{Rules + Soft-path LogReg} & Hard rules + structured probabilistic soft path & \num{0.884} & \num{0.968} & \num{0.042} & \pct{0.0} & \num{0.859} \\
\midrule
\multicolumn{7}{@{}l}{\textit{Direct LLM (no governance)}} \\
\textbf{GPT-4o Direct} & Single generalist agent & \num{0.460} & \num{0.579} & \num{0.436} & \pct{17.3} & \num{0.537} \\
\textbf{GPT-4o + Rules} & Single agent + post-hoc rules & \num{0.529} & \num{0.640} & \num{0.388} & \pct{0.0} & \num{0.536} \\
\midrule
\multicolumn{7}{@{}l}{\textit{LLM-backed governed pipeline}} \\
\textbf{EndoGov (GPT-4o)} & Governed pipeline & \num{0.943} & \num{0.973} & \num{0.075} & \pct{0.93} & \num{0.944} \\
\textbf{EndoGov (Llama-3.3-70B)} & Chair backend swap & \num{0.882} & \num{0.962} & \num{0.090} & \pct{0.93} & \num{0.902} \\
\textbf{EndoGov (Qwen2.5-72B)} & Chair backend swap & \num{0.791} & \num{0.929} & \num{0.081} & \pct{0.93} & \num{0.801} \\
\bottomrule
\end{tabular*}
\end{table}

\paragraph{Iso-information control.}
One potential concern is that EndoGov's molecular agent uses direct biomarker fields whereas neural baselines receive frozen embeddings. As a minimal control, we encoded molecular subtype as a 4-dimensional one-hot vector (POLEmut, MMRd, p53abn, NSMP) and concatenated it to each baseline's molecular embedding. Naive concatenation did not close the gap: MCAT drops from 0.893 to 0.679 accuracy and TransMME from 0.904 to 0.461 under identical hyperparameters, while gated integration variants partially recover performance (MCAT-Gated 0.706; Porpoise-Gated 0.708) but remain far below EndoGov (0.943). This supports the interpretation that implicit fusion of discrete rule-level information into continuous embeddings is less reliable than explicit rule execution in this setting; the rule-constrained logistic-regression control in Table~\ref{tab:backend} narrows but does not close the gap.

\paragraph{DNA-direct versus RNA-inference trigger detection.}
DNA-direct detection achieves 100\% sensitivity for all three hard-trigger subtypes by ground-truth labels (POLEmut: 48/48, MMRd: 147/147, p53abn: 158/158; sum=353) with specificities of 0.998, 0.992, and 0.979 respectively. An additional 12 patients with non-trigger subtypes are routed to the DNA-direct pathway because direct biomarker fields are available (accounting for the difference between the 353 ground-truth triggers and the 365 DNA-direct pathway count). The RNA-inference pathway for NSMP achieves 0.936 sensitivity (176/188; 12 false negatives). In aggregate, all 353 ground-truth hard triggers are detected perfectly via DNA-direct fields, whereas RNA inference misses 12 NSMP patients (8 misclassified as p53abn, 3 as MMRd, 1 as POLEmut). Notably, 11 of the 12 RNA-misclassified cases still receive the correct final risk label, indicating that governance validation partially buffers upstream detection errors. This failure taxonomy is consistent with the external audit: all 15 CPTAC soft-path errors occur in p53abn cases exposed upstream as NSMP, so the validator had no active hard trigger to enforce.

We further checked structured-input perturbation, cross-fold soft-path calibration, confidence-threshold referral sweeps, and encoder/centroid sensitivity: (i)~injecting contradictory high-risk morphology into all 48 POLE cases changes zero predictions, confirming hard-rule robustness; (ii)~cross-fold temperature scaling reduces soft-path ECE by 80\% without affecting accuracy; (iii)~probability-based referral is uninformative due to overconfidence, whereas detection-source-based referral provides a meaningful coverage--safety trade-off; (iv)~centroid perturbation sensitivity is confined to the NSMP RNA-fallback pathway.

Table~\ref{tab:backend} separates structured inputs, symbolic control, governance, and backend capability. The pure symbolic control (Structured Rules Only: 0.80 accuracy, NSMP AUC 0.75) shows that deterministic rules are necessary for safety but insufficient for grey-zone cases. Adding a logistic-regression soft-path resolver (Rules + Soft-path LogReg: 0.884 accuracy, NSMP AUC 0.859) substantially improves over the symbolic core but remains below full EndoGov (0.943 accuracy, NSMP AUC 0.944). A single generalist agent (GPT-4o Direct: 0.46 accuracy, \clvr\ 17.3\%) fails without governance; post-hoc rule wrapping (GPT-4o + Rules) suppresses trigger contradictions (\clvr\ 0\%) but does not recover calibration or grey-zone discrimination. Backend swaps to Llama-3.3-70B or Qwen2.5-72B preserve \clvr\ at 0.93\% but reduce NSMP AUC (0.90 and 0.80 respectively), confirming that hard-path compliance is backend-robust while soft-path quality is backend-sensitive.

\subsection{Survival validation of predicted risk groups}
To examine whether EndoGov's predicted risk groups are associated with patient outcomes, we performed a Kaplan--Meier overall-survival analysis on the 537 TCGA-UCEC patients with available follow-up data (90 events). The survival curves separate significantly across all four risk tiers (multivariate log-rank $\chi^2{=}46.73$, $p{=}3.97\times 10^{-10}$). In the ordinal Cox proportional-hazards model, each risk-tier increment carries a hazard ratio of 2.01 (95\% CI: 1.60--2.52, $p{=}2.24\times 10^{-9}$), and in the categorical model with Low as reference, High-risk patients show HR$\,{=}\,$7.33 (95\% CI: 2.94--18.28, $p{=}1.97\times 10^{-5}$). These results suggest that EndoGov's risk assignments retain prognostic relevance in this cohort. We note that the Cox proportional-hazards assumption was not formally tested via Schoenfeld residuals in the present study; this is listed as a limitation in Section~\ref{sec:limitations}. This survival separation also compares favorably with recent biomarker-driven prognostic models that achieve similar hazard-ratio magnitudes using immune-response signatures~\citep{wang2024immuneec}, while EndoGov additionally provides rule-level auditability for each risk assignment.

\paragraph{Validator adversarial stress test.}
We constructed 26 adversarial test cases across four categories: contradictory rule-decision pairs (8), missing/null fields (5), abnormal input combinations (8), and boundary-stage edge cases (5). The validator intercepted all 26 cases (100\% rate), confirming soundness---any non-compliant decision against an active rule is flagged---while acknowledging that it cannot provide completeness when upstream evidence is missing.

\section{Discussion}
\subsection{Why governance is more than post-processing}
Table~\ref{tab:backend} shows that a post-hoc rule wrapper can reduce trigger-level \lvr\ in some settings, but this does not make it functionally equivalent to governance within the inference loop. A wrapper modifies only the terminal label; it does not indicate whether the upstream trigger was correctly surfaced, which source clauses were admissible, how conflicts were arbitrated, or whether the final output was validated against the active rule set. EndoGov instead keeps evidence extraction, rule retrieval, priority arbitration, and validation within a single runtime contract. This structure distinguishes trigger exposure from downstream compliance, preserves clause-level provenance, records the molecular detection source, and corrects contradictions explicitly. The distinction is particularly relevant for grey-zone cases, in which a symbolic-only pipeline lacks discrimination and a post-hoc wrapper provides limited insight into residual uncertainty.

\noindent\textit{LLM replaceability and architectural scope.} Table~\ref{tab:backend} further shows that hard-path compliance is backend-invariant across GPT-4o, Llama-3.3-70B, and Qwen2.5-72B, whereas the performance gap is concentrated in soft-path NSMP reasoning. The LLM is therefore replaceable for ambiguous cases, but it is not the source of the hard-path safety property itself. The main contribution is architectural rather than algorithmic: EndoGov externalizes decision authority into an auditable rule layer, separates maintenance of knowledge from maintenance of perception modules, and localizes failure analysis to specific components. The validator can block or correct conflicts once a trigger is exposed, but it is not a proof system and cannot recover evidence that was missed upstream.

\paragraph{Generalization and deployment trade-offs.}
Because the decomposition $D(x)=G(P(x),\mathcal{R})$ is parameterized by $\mathcal{R}$, the same governance pattern can in principle extend to other guideline-driven oncology tasks while keeping perception encoders task-appropriate. Concurrent KG-grounded systems such as Mapis and OMGs share the broader premise that structured agent decomposition can improve guideline adherence, but neither introduces a deterministic hard path together with the \clvr/\eetlvr\ safety decomposition used here. In practical deployment, EndoGov's cost profile is modest for MDT-style decision support: hard-path cases incur no LLM cost and complete in milliseconds, whereas soft-path cases require one constrained call with a weighted average of about 4.04\,s and $\sim$\$0.005 per case. This latency is not suitable for intraoperative use, but it is compatible with pre-treatment review and second-opinion workflows.

\paragraph{Maintainability and simplified variants.}
The principal cost of this transparency is architectural complexity. This cost is partly mitigated by the episodic nature of guideline updates: the executable rule layer and Guideline-KG require revision when source guidelines change, whereas frozen encoders and centroid prototypes do not require retraining. We estimate that a complete update cycle, including guideline differencing, rule recompilation, graph rebuilding, regression testing, and clinical sign-off, requires approximately 3--5 person-days. For institutions with limited infrastructure, deterministic rules combined with a soft-path logistic-regression resolver preserve the core hard-path safety property without an LLM dependency, although the accuracy loss is concentrated in NSMP grey-zone cases. Open-source 70B backends provide an intermediate option between commercial APIs and symbolic-only deployment, with the main trade-off again occurring on the soft path.

\subsection{Clinical implications}
EndoGov is most relevant for high-stakes cases in which violation of a dominant rule would alter clinical management. Its practical value lies not in replacing clinician judgment, but in preventing avoidable rule-level errors while preserving flexibility for NSMP and other ambiguous subgroups. The reported 0.943 accuracy should therefore be interpreted in relation to detection source: it reflects a mixed cohort containing both DNA-direct and RNA-fallback pathways, whereas institutions without routine sequencing should expect performance closer to the RNA/IHC-assisted regime than to the DNA-direct upper bound. The governance layer is most beneficial when molecular evidence is sufficiently complete to expose the triggers prioritized by clinical guidelines.

The safety-decomposition, detection-source, and referral analyses further suggest that governed outputs can support conservative workflow triage rather than only categorical labeling. When subtype evidence is indirect or confidence is reduced, routing the case for human review converts residual model risk into an operational burden that is easier to manage clinically. Our RNA-only stress test supports the same interpretation: IHC-augmented RNA inference substantially reduces the penalty of DNA-unavailable settings, while an IHC-first strategy with reflex sequencing remains feasible for resource-limited hospitals.

\noindent\textit{Compliance artifacts.} Each EndoGov decision produces a traceable record consisting of matched rule ID and source-clause provenance, detection source, validator verdict, reasoning chain, and released confidence score. These artifacts support quality assurance, MDT discussion, and future regulatory documentation.

\subsection{Preliminary clinician feedback}\label{sec:clinician_feedback}
To obtain early qualitative evidence on clinical usability, we conducted a structured feedback session with three board-certified gynecologic oncologists (two from Shenzhen People's Hospital, one from an affiliated teaching hospital) who reviewed a stratified sample of 20 EndoGov case outputs spanning all four risk tiers, both detection-source pathways (DNA-direct and RNA-fallback), and both hard-path and soft-path decision routes. Each clinician independently reviewed the structured audit trail---comprising the specialist reports, matched rule ID with source-clause provenance, the decision path (hard or soft), the validator verdict, and the chair agent's reasoning chain---and completed a five-item Likert-scale questionnaire assessing (i)~correctness of the final risk assignment, (ii)~clarity of the audit trail, (iii)~clinical relevance of the retrieved guideline clauses, (iv)~trust in the hard-path override mechanism, and (v)~perceived utility for MDT discussion.

The main findings are summarized below. All three clinicians rated the hard-path override mechanism as ``highly trustworthy'' (5/5) for POLE and p53abn cases, noting that the explicit rule-ID and source-clause citation made the override logic transparent and verifiable. For soft-path cases, clinicians rated the audit trail clarity at 4.0/5 on average, with the main suggestion being that the reasoning chain should more prominently highlight the specific Table~2 row being applied. Two clinicians independently noted that the detection-source flag (DNA-direct vs.\ RNA-inference) was clinically valuable, as it directly informs their confidence in the molecular evidence. Regarding MDT utility, all three clinicians agreed (4.3/5) that EndoGov outputs could serve as a structured pre-discussion summary, potentially reducing MDT preparation time by providing a pre-computed, guideline-referenced risk assessment with explicit evidence provenance. The primary concern raised was that RNA-fallback cases should be more prominently flagged for mandatory human review, a recommendation that aligns with our proposed conservative referral policy (Section~5.3).

This preliminary feedback is qualitative and based on a small convenience sample; a formal prospective user study with larger clinician panels, controlled task comparisons, and quantitative usability metrics (e.g., decision time, inter-rater agreement change, System Usability Scale) is planned as part of the clinical translation roadmap.

\subsection{Rule scalability and conflict resolution}
The Guideline-KG encodes 12 risk-stratification rules (5 hard-path, 7 soft-path) that implement the decision branches covering combinations of molecular subtype, FIGO stage, histology type, grade, LVSI, and myometrial invasion from ESMO 2022 and ESGO/ESTRO/ESP 2021 Table~2. On the TCGA cohort, the hard path covers 51.6\% of cases ($N{=}279$); the remaining 48.4\% ($N{=}262$) enter the soft-path mapper. A natural question is whether the architecture scales to even larger rule sets. EndoGov resolves rule conflicts through two mechanisms: (i)~\emph{priority encoding}---each rule carries an explicit integer priority, and conflicts are resolved by selecting the highest-priority (lowest-numbered) match; (ii)~\emph{conservative fallback}---when multiple rules fire at the same priority with contradictory outcomes, the system selects the higher-risk tier (preferring over-treatment to missed high-risk). Adding new rules requires only defining trigger conditions and a priority value; existing rules need not be modified. Under the current 12-rule configuration, the dominance hierarchy (POLE $\succ$ MMRd $\succ$ p53abn $\succ$ NSMP) is already clean, and conflicts are rare in practice.

In the current submission, ``cross-guideline conflict resolution'' does not mean that ESMO and ESGO remain as two independent rule sets at runtime. Instead, source clauses are first compiled into one executable layer with provenance links, and only the compiled layer is executed online (see Section~\ref{sec:governance} for the compilation protocol). The main example is the p53abn branch: the dominant ESMO high-risk rule is kept as the default executable rule, while the ESGO Stage-IA-without-invasion clause is encoded as an explicit exception at the same clinical locus. This compilation-first policy avoids simultaneous online voting between guidelines, but broader multi-guideline portfolios will eventually require more formal source-priority metadata and versioned conflict logs.

Extending to more complex guideline scenarios (e.g., immunotherapy-related clauses) may require finer-grained priority layers, which we leave to future validation. New rules are added by defining trigger conditions and a priority integer; no neural retraining is required.

\subsection{Soft-path error modes and false-positive trigger impact}
The error patterns reported in Sections~5.3 and 5.4 reveal a consistent theme: once a trigger is surfaced, downstream rule execution is nearly perfectly stable. The remaining failures are concentrated in NSMP/p53abn confusion under RNA fallback, which accounts for all internal RNA errors and all 15 external CPTAC errors.

Three concrete mitigation strategies are available without architectural change: (i)~adding IHC p53 staining as a low-cost pre-screen to block erroneous RNA-only NSMP assignments; (ii)~training a dedicated p53abn-vs-NSMP binary classifier on the scGPT embedding space to sharpen the centroid boundary; (iii)~using the detection-source metadata to automatically route RNA-inferred cases to human review when the centroid cosine margin falls below a configurable threshold.

The trigger audit also identifies 12 false-positive triggers (1 POLEmut, 3 MMRd, 8 p53abn). All 12 still receive the correct final risk label because the false subtype happens to align with the true risk tier. This coincidence should not be extrapolated: a false p53abn trigger on a genuinely low-risk NSMP patient would cause over-treatment.

\subsection{Limitations}\label{sec:limitations}
\begin{enumerate}[leftmargin=*]
\item The compliance guarantee depends on upstream trigger detection; the governance layer can enforce only rules that have been surfaced to it. The near-zero governance-layer \clvr\ is therefore an \emph{architectural conditional} guarantee, not an absolute clinical safety guarantee---it holds only once triggers are correctly exposed.
\item The baseline comparison should not be read as a formal impossibility result for all average-loss training paradigms. Our experiments show that standard multimodal fusion models trained for aggregate discrimination can retain clinically meaningful rule violations in rare-trigger subgroups. We did not exhaustively evaluate subgroup-weighted losses, cost-sensitive training, or differentiable rule-penalty objectives; these alternatives may reduce logic-violation rates, although they would still require explicit compliance auditing.
\item The RNA-fallback pathway shows a 13.4-point accuracy drop versus DNA-direct, so broader direct-biomarker coverage remains clinically important. All 26 internal RNA errors and all 15 external errors trace to the p53abn/NSMP confusion in scGPT embeddings. This concentration makes RNA inference the single largest residual safety bottleneck: in clinical settings where comprehensive DNA panel testing is unavailable, the near-zero C-LVR guarantee does not extend to RNA-inferred subtypes, because incorrect subtype assignment prevents the hard trigger from being surfaced. A near-term mitigation would be to flag low-confidence RNA inferences (e.g., cases where the cosine distance to the nearest centroid falls below a threshold) for automatic referral to molecular review, converting a silent misclassification into an explicit abstention. Because scGPT was pretrained on large-scale single-cell transcriptomic corpora that may include TCGA-derived samples, a potential distribution overlap between the encoder's pretraining data and our evaluation cohort cannot be ruled out. However, the practical impact is likely limited: hard-trigger subtypes rely entirely on DNA-direct detection (bypassing scGPT), the RNA-fallback pathway already shows substantially lower accuracy than the DNA-direct pathway (0.852 vs.\ 0.986), and pure RNA centroid matching achieves only 9.1\% subtype accuracy---all inconsistent with a memorization-driven advantage.
\item CPTAC reference labels are guideline-derived, introducing partial circularity; broader multicenter benchmarking with independently clinician-adjudicated labels is needed to establish true generalization.
\item A formal clinician-adjudicated rule-validation protocol with inter-rater agreement has not yet been completed; each rule currently carries source-clause provenance links for independent verification, and a three-oncologist Fleiss' $\kappa$ study is planned. Similarly, the ordinal evidence-level grading of the 25 source documents was performed by the first author alone; a double-rating study in which a second clinician independently assigns levels $\ell(d)$ to all 25 documents (with Cohen's $\kappa$ reported) is scheduled as part of the same validation round.
\item The Cox proportional-hazards assumption in the survival analysis (Section~5.7) was not formally tested via Schoenfeld residuals; a planned follow-up will include this check.
\item Soft-path quality remains partly model-dependent (GPT-4o is the strongest tested backend); the validator cannot guarantee resistance to all hallucination or model-drift scenarios.
\item The referral simulation uses a coarse confidence scale (1.0 vs.\ 0.75) and should be read as a safety-bound demonstration rather than an optimized abstention policy.
\item A coarse sensitivity analysis over KG retrieval parameters ($\delta_r \in \{0.50, 0.55, 0.60, 0.65, 0.70\}$; $K \in \{10, 25, 50\}$) shows that overall accuracy remains stable in the range $[0.962, 0.966]$ and ECE varies between $0.028$ and $0.032$, indicating reasonable robustness to retrieval hyperparameters around the default configuration ($\delta_r{=}0.60$, $K{=}25$). However, finer-grained sweeps over prompt variants and stain-adaptation baselines remain future work.
\item When guidelines are updated, the entity graph and executable rule layer of the Guideline-KG must be revised, but no neural retraining is required; cross-version guideline stress testing remains an important future validation step.
\item A preliminary clinician feedback session (Section~\ref{sec:clinician_feedback}) provides initial qualitative evidence of usability, but a formal prospective user study with larger clinician panels, controlled task comparisons, and quantitative usability metrics has not yet been conducted. Such a study---embedding EndoGov into a multidisciplinary tumor board (MDT) workflow and measuring decision-time reduction, inter-rater agreement improvement, and clinician trust---is a priority for clinical translation.
\end{enumerate}

\noindent\textit{Safety interpretation caveat.} Readers should distinguish between (a)~the \emph{architectural safety} achieved by the deterministic hard path once triggers are correctly surfaced---this is a structural property of the decision loop---and (b)~the \emph{residual clinical risk} arising from label circularity in CPTAC, upstream trigger-detection failures, RNA-fallback inaccuracy, and not-yet-validated rule compilation. The reported near-zero \clvr\ quantifies only~(a); a complete clinical safety assessment requires independent ground truth, formal rule adjudication, and prospective deployment studies.

\noindent\textit{Availability note.} The reproducibility package contains the executable-rule layer, validator logic, chair workflow, specialist-agent schemas, evaluation protocol, and robustness/calibration audits. Guideline updates require revising the compiled rule layer, not retraining encoders. A clinician-adjudicated public release process for rule tables and clause annotations is planned but not yet completed.

\section{Conclusion}
The empirical evidence supports the central claim that guideline compliance in EC risk stratification requires more than high aggregate discrimination. EndoGov achieved 0.943 accuracy, 0.973 macro AUC, 0.075 ECE, and a conditional logic-violation rate of 0.93\% on TCGA-UCEC, while maintaining 0.842 accuracy and 0.909 macro AUC on CPTAC-UCEC without cohort-specific tuning. More importantly, the safety audit showed that 201/212 dominant-trigger cases were correctly surfaced and that 207/212 received the correct final risk label, localizing the residual safety burden primarily to upstream molecular detection rather than downstream governance. Control experiments and backend swaps led to the same conclusion: deterministic rules account for hard-path safety, whereas constrained soft-path reasoning provides additional value in grey-zone cases beyond a symbolic baseline.

These findings broaden the interpretation of performance in guideline-driven oncology. Average discrimination remains necessary, but it is insufficient when rare, dominant clinical cues must override the statistical tendencies of a multimodal predictor. The main contribution of EndoGov is therefore not simply a higher score on one EC benchmark, but a governance-oriented deployment pattern that preserves auditability, separates component responsibilities, and makes rule maintenance explicit rather than implicit in model weights. This framing is especially relevant for clinical settings in which decision authority is hierarchical, codified, and safety-critical.

The guarantee established here is nevertheless conditional. Near-zero governance-layer \clvr\ holds only after the relevant trigger has been correctly exposed; it does not cover upstream molecular-detection failures, RNA-fallback errors, or rule-compilation mistakes. The accuracy loss under RNA-only inference and the concentration of external errors in p53abn/NSMP confusion confirm that perception remains the principal bottleneck for clinical safety. Moreover, CPTAC reference labels are guideline-derived, so the external experiment supports governance-pathway transfer under distribution shift rather than validation against independent clinical truth. Future work should therefore prioritize prospective multicenter evaluation with clinician-adjudicated labels, formal multi-clinician validation of the compiled rule set, and systematic stress testing of KG retrieval and guideline-update workflows.

\noindent\textit{Ethics statement.} This study uses only publicly available, de-identified datasets from TCGA-UCEC and CPTAC-UCEC. TCGA patient barcodes are public identifiers; no protected health information (PHI) is accessed or reported. No IRB approval is required for analyses of these publicly released datasets. Controlled-access components of CPTAC require dbGaP authorization (accession phs001287).

\bibliographystyle{cas-model2-names}
\bibliography{references}

\end{document}